\documentclass[fleqn,usenatbib,useAMS,twocolumn]{mnras}

\usepackage[T1]{fontenc}

\DeclareRobustCommand{\VAN}[3]{#2}
\let\VANthebibliography\thebibliography
\def\thebibliography{\DeclareRobustCommand{\VAN}[3]{##3}\VANthebibliography}

\usepackage{graphicx}	
\usepackage{amsmath}	
\usepackage{amssymb}	
\usepackage{txfonts}

\title[Cosmological evolution of halo gas and SMBHs]{Cosmological evolution of gas and supermassive black holes in idealized isolated halos}

\author[Shashank Dattathri, Prateek Sharma]{
Shashank Dattathri,$^{1,2}$\thanks{E-mail: shashank.dattathri@gmail.com}
Prateek Sharma,$^{1}$
\\
$^{1}$Department of Physics, Indian Institute of Science, Bangalore-560012, India\\
$^{2}$Department of Astronomy, University of Michigan, 1085 S. University Avenue, Ann Arbor, MI, 48109, USA
}

\date{Accepted 2022 May 4; received 2022 April 10; in original form 2021 November 7}

\defcitealias{prasad15}{P15}
\defcitealias{dk14}{DK14}

\pubyear{2021}

\begin{document}
\label{firstpage}
\pagerange{\pageref{firstpage}--\pageref{lastpage}}
\maketitle

\begin{abstract}
We study the evolution of baryonic gas in cosmologically growing dark matter halos. To accurately model both the inner and outer regions of the halos, we use a dark matter density profile that transitions smoothly from the NFW profile within the virial radius to a more realistic flat profile far beyond the halo. We construct a dark matter gravitational potential consistent with this density profile, and we use a "cosmological" potential that accounts for gas evolution consistent with Hubble expansion at large radii. Gas is initialized with a density $\approx$ 0.2 times the dark matter density, consistent with the universal baryon fraction $\rho_{\rm g}/(\rho_{\rm g}+\rho_{\rm DM}) \approx 0.17$. We study the formation of the virial shock and evolution of the baryon fraction, including the effects of radiative cooling and AGN jet feedback. The feedback is powered by the accretion of cold gas onto a central supermassive black hole (SMBH). The cores of the halo exhibit heating and cooling cycles, whose strength and duration depend on the feedback efficiency and the halo mass. The central SMBH initially grows exponentially with time in the early quasar phase, but the growth slows down at later times. The baryon fraction in the core decreases with increasing feedback efficiency and decreasing halo mass. While the halo outskirts evolve self-similarly, the core density is non-evolving, in agreement with cluster observations. We analyze the correlations between the properties of the gas and the central SMBH, and explore the existence of a fundamental plane.
\end{abstract}

\begin{keywords}
Galaxy: active--Galaxies: clusters: intracluster medium--Galaxies: halos
\end{keywords}



\section{Introduction}
According to the standard $\Lambda$CDM ($\Lambda$ cold dark matter) paradigm of structure formation in the Universe, the evolution of galaxy clusters is primarily governed by dark matter halo dynamics. Whereas, in addition to following the dark matter gravity, the gas (which constitutes $\sim$ 80\% of baryons within massive halos) is strongly affected by radiative cooling and feedback heating powered by accretion onto a central supermassive black hole (SMBH). Cosmological N-body simulations provide a good description for the large-scale (on scales larger than the halo) distribution of baryonic and dark matter in the Universe. For example, the Millennium simulation \citep{millenium} evolves over $10^{10}$ particles and provides predictions for structure formation in the Universe in broad agreement with theoretical and observational works. However, the evolution of baryons within individual halos is more complex and requires including processes such as radiative cooling and feedback heating. Modern cosmological galaxy formation simulations, such as EAGLE \citep{eagle}, Magneticum \citep{magneticum} and IllustrisTNG \citep{illustristng}, evolve baryons including these processes, and with sub-grid models for star formation and SMBH growth (see \citealt{vogelsberger2020} for a review). 

Another class of numerical simulations consists of idealized halo simulations that focus on various aspects of baryonic physics. While these simulations  provide insight into gas evolution, they typically lack cosmological evolution. Most of the idealized simulations of isolated halos assume a static dark matter halo (e.g. \citealt{prasad15} [hereafter \citetalias{prasad15}], \citealt{fielding}), and various important parameters (such as the metallicity of the IGM, e.g. \citealt{choudhury}) are not evolved cosmologically. While modern large-scale cosmological galaxy formation simulations reach very high resolutions (spatial and mass resolutions $\sim$ 1 kpc and $\sim 10^7 M_\odot$ respectively), they cannot achieve the resolution achievable in single-halo or zoom-in simulations (spatial and mass resolutions $\sim$ 0.01 kpc and $\sim 10^4 M_\odot$ respectively). More importantly, because cosmological galaxy formation simulations model several  unresolved/sub-grid physical processes (star formation, black hole accretion, etc.), they contain a large number of free parameters that must be fine-tuned to match observations. The free parameters may lead to degeneracies, making it difficult to establish causality between various processes/parameters and the observables. Therefore, we focus on the most basic physical processes governing the halo gas and  carry out simulations of the halo gas in cosmologically growing halos. This approach is numerially less expensive and can provide a useful middle ground between fully cosmological and isolated, cosmologically non-evolving halo simulations.

Dark matter halos grow hierarchically, through the accretion of matter from the surrounding medium and through mergers with other halos. Since this growth is non-linear, semi-analytical models of the formation and evolution of dark matter halos are usually based on cosmological N-body simulations. The standard spherical infall model of matter onto dark matter halos predicts a density profile $\rho \propto r^{-2}$ (e.g., see \citealt{mo2010}). This is in rough agreement with N-body simulations. The density profile of dark matter within halos in cosmological N-body simulations is well described by the NFW (Navarro-Frenk-White) profile \citep{nfw}. However, the outer regions of the halo do not follow this profile as the density cannot continuously decrease with radius and must match the average density of the Universe far away from the halo. \citet{dk14} (hereafter \citetalias{dk14}) study the dependence of the outer density profile of a halo on its mass accretion rate and give a more realistic profile that applies outside of the virial radius. Thus, it is necessary to accurately model both the inner and outer regions of the halo since over time the gas/dark matter in halo outskirts is accreted within.

In the absence of radiative cooling and feedback heating, the gas is expected to roughly follow the dark matter profile and behave self-similarly. The gas within the halo thermalizes via an outward moving pressure-supported virial shock \citep{dekel}. The innermost regions of the halo form a core with a nearly constant density and temperature. The gas evolution closely follows the dark matter  \citep{choudhury}, leading to a baryon fraction within the virial radius close to the universal value of $1/6$ \citep{wmap}.

However, the hot gas profiles as measured by X-ray observations are not self-similar in groups and clusters \citep{ponman2003}. Since cooling efficiency of hot gas depends strongly on the temperature, radiative cooling is expected to break self-similarity in baryons. The cooling time in the inner regions of the halo is short, which in absence of heating would lead to catastrophic cooling and intense star formation. This cooling flow model predicts a gas inflow rate of $100-1000 \ M_\odot \rm{yr}^{-1}$ in cool core clusters, which is typically at least an order of magnitude greater than what is observed. Thus, some source(s) of heating is required in order to prevent catastrophic cooling and the associated star formation. 

A promising heating mechanism is feedback due to active galactic nuclei (AGN) powered by accretion onto the central supermassive black hole (SMBH). Although the SMBH is dynamically unimportant over $\gtrsim$ kpc scales, it can inject sufficient energy in the halo gas to control gas cooling in halos at much larger scales, and hence regulates its own growth and the growth of the central galaxy. Several observational studies by Chandra and XMM-Newton (e.g., \citealt{bohringer2002,birzan2004}) have shown the presence of X-ray cavities (also seen as radio bubbles) which can be attributed to AGN jets. Numerous studies have shown that the AGN feedback can suppress the mass inflow rate within cluster cores by about an order of magnitude compared to a cooling flow (\citealt{gaspari2012,li2014a}; \citetalias{prasad15}). The accretion rate onto the black hole depends on this mass inflow rate, the efficiency of mass transport from $\sim 1$ kpc to the event horizon ($\lesssim 10^{-3}$ pc), and is expected to be limited by the Eddington rate. The evolution of the central supermassive black hole is characterized by an initial phase where the accretion rate is high (termed as the quasar regime) and a later phase where the accretion rate is low (the radio regime). This has important implications for the growth of SMBHs \citep{sijacki}, as most rapid growth of SMBHs occurs during the quasar phase. In addition, the Eddington rate limits the feedback energy that can be injected into the ICM at high redshift, which prevents the gas from being completely ejected from the core.

In this work, we study the evolution of baryonic gas in cosmologically evolving dark matter halos, incorporating various aspects of cosmology as well as baryonic physics. We evolve the dark matter halo mass and density profiles, and IGM/halo metallicities according to prescriptions calibrated with cosmological simulations. We monitor the central supermassive black hole that grows because of gas cooling at $\sim 1$ kpc. The regions beyond the halo evolve according to Hubble expansion. Our model is therefore self-consistent with basic aspects of cosmological evolution and structure formation. We do not include halo or black hole mergers in our model, but instead choose to focus on the self-regulated feedback induced by the AGN in relaxed/quiescent halos. 

This paper is organized as follows. In section \ref{sec:setup}, we describe our semi-cosmological setup for studying the baryonic gas in an evolving halo. We describe the results of our simulations in absence of radiative cooling and feedback, with only cooling, and with cooling and AGN jet feedback in sections \ref{sec:nonrad}, \ref{sec:cooling_runs}, and \ref{sec:AGN} respectively. We discuss our results and their astrophysical implications in section \ref{sec:discuss}, and conclude in section \ref{sec:conc}.

\section{Physical setup} \label{sec:setup}
We use spherical coordinates ($r$, $\theta$, $\phi$) to solve the standard hydrodynamical equations with external gravity, radiative cooling, and mass and momentum injection due to AGN jet feedback. Since we explore a large parameter space, we only carry out 1-D and 2-D $(r,\theta)$ axisymmetric simulations. In conservative form, the equations are given by 
\begin{equation} 
\label{eqt_density}
\frac{\partial \rho}{\partial t}+ \nabla \cdot (\rho \boldsymbol{v})= S_\rho \; ,
\end{equation}
\begin{equation}
\frac{\partial (\rho \boldsymbol{v})}{\partial t}+\nabla \cdot (\rho \boldsymbol{v} \boldsymbol{v}) =- \nabla P -\rho \nabla \Phi + S_\rho \boldsymbol{v}_{\rm jet} \; ,
\end{equation}
\begin{equation}
\begin{split}
&\frac{\partial}{\partial t}\left(\frac{\rho v^2}{2}+\frac{P}{\gamma-1} \right) + \nabla \cdot \left[ \left(\frac{\rho v^2}{2}+\frac{\gamma P}{\gamma-1} \right) \boldsymbol{v} \right] = S_\rho \left( \boldsymbol{v}_{\rm jet} \cdot \boldsymbol{v} -\frac{v^2}{2} \right) \\
&\quad +S_e-\rho \boldsymbol{v} \cdot \nabla \Phi-\mathcal{L}_{\rm cool} \; ,
\end{split}
\end{equation}
where $\mathcal{L}_{\rm cool}=n_e n_i \Lambda$ is the energy loss rate per unit volume due to radiative cooling, and the AGN terms $\boldsymbol{v}_{\rm jet}$, $S_\rho$, and $S_e$ refer to the AGN jet velocity, mass injection, and thermal energy source terms. We note that the kinetic energy and thermal energy injection terms due to the jet separate out in the energy equation. We solve these equations using the astrophysical gas dynamics code PLUTO \citep{pluto}. PLUTO is a conservative Godunov (magneto)hydrodynamic code which incorporates several different modules such as tabulated radiative cooling. We solve Euler equations (with numerical dissipation) in presence of gravity, radiative cooling and feedback heating using a static spherical ($r,\theta,\phi$) grid. The time-stepping is performed using the explicit $3^{rd}$ order Runge-Kutta algorithm, and we use the Harten, Lax, van Leer (HLL) Riemann solver for flux computation.
We use the cosmological parameters $\Omega_m=0.27$, $h=0.7$, $\sigma_8=0.82$, and $n_s=0.95$ \citep{wmap}. 
\subsection{Dark matter halo}
The NFW profile \citep{nfw} 
does not accurately model the outer regions of the halo, beyond which the matter density should asymptote to the mean density of the Universe. We therefore initialize the dark matter density profile following \citetalias{dk14} that transitions from a NFW profile in the inner regions to a realistic outer profile:
\begin{equation}
\label{eq:dk14first}
\rho_{\rm in}(r)=\rho_{\rm NFW}(r)=\frac{\delta_c \rho_{\rm cr}}{r/R_s \left( 1+r/R_s \right)^2},
\end{equation}
\begin{equation}
\label{eq:dk14_out}
\rho_{\rm out}(r)=\rho_m \left[ b_e \left(\frac{r}{5 R_{200m}} \right)^{-s_e} +1 \right],
\end{equation}
\begin{equation}
\label{eq:ftrans}
f_{\rm trans}=\left[ 1+\left( \frac{r}{r_t} \right)^\beta \right]^{-\frac{\gamma}{\beta} },
\end{equation}
\begin{equation}
\label{eq:dk14last}
\rho(r)=\rho_{\rm NFW} f_{\rm trans}(r)+\rho_{\rm out},
\end{equation}
where $$\delta_c = \frac{200}{3} \frac{c^3}{[\ln(1+c)-c/(1+c)]},$$ 
$\rho_{\rm cr} = 3 H^2/8 \pi G$ is the critical density of the Universe, $R_{200m}$ ($R_{200c}$) is the radius within which the average matter density is 200 times the mean (critical) density of the Universe, $R_S=R_{200c}/c$ ($c$ is the concentration parameter), $b_e$ and $s_e$ characterize the dark matter density in the outer regions of the halo, and $\beta$, $\gamma$ and $r_t=1.18 R_{200m}$ characterize the transition factor. The dependence of these parameters on the various halo properties is examined in \citetalias{dk14}. At all times during the simulation, the DM density profile follows equation \ref{eq:dk14last}, with the characteristic radii ($R_{200c}$ and $R_{200m}$ and, in some cases, other parameters) evolving over time. Our test simulations with just the NFW potential extended to large radii ($\rho(r) = \rho_{\rm NFW}(r)$) confirm that the gravity in halo outskirts is weak and the gas density at the virial radius and beyond is smaller by $\sim 2$. 
\subsection{Evolution of the halo}
\begin{figure}
	\includegraphics[width=\columnwidth]{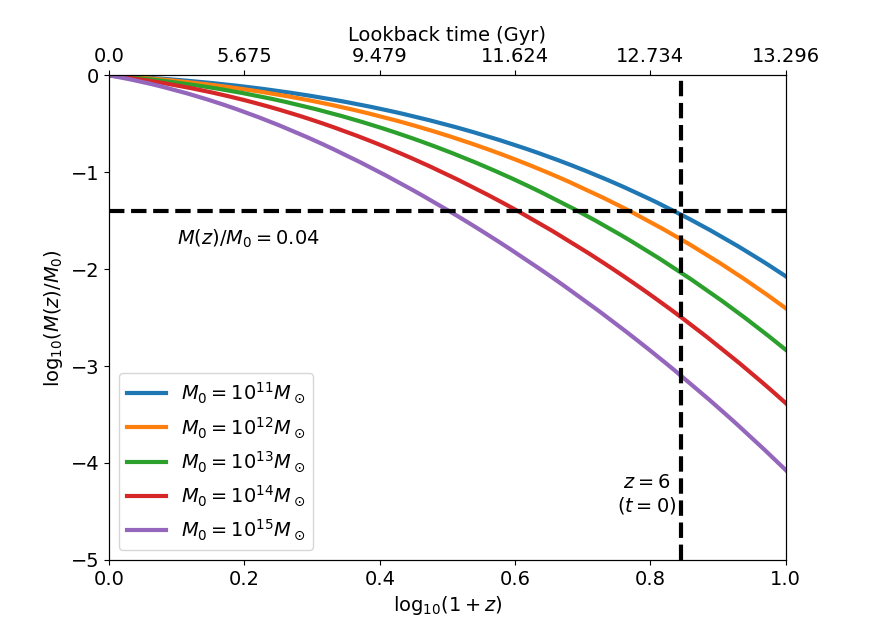}
	\caption{Mass accretion histories of dark matter halos with the $z=0$ mass ranging from $10^{11}$ to $10^{15} M_\odot$. These growth histories are calculated according to \citet{vdb14_1}, using the cosmological parameters $\Omega_m=0.27$, $h=0.7$, $\sigma_8=0.82$, and $n_s=0.95$ \citep{wmap}. We note that more massive halos grow faster than less massive ones and therefore have smaller values of $t_{0.04}$, the time when they reach 4\% of the mass at $z=0$ (see equation \ref{conc}). We begin our runs at $z=6$, which corresponds to $t=0$ in our simulations. 
	} 
	\label{fig:halo_acc}
\end{figure}

In our work, we are not concerned with the detailed distribution of the masses of the progenitor halos or the exact rate of mergers. Rather, we wish to model the average mass of the main halo as a function of time. Although the accretion history of each individual halo is unique, the average mass accretion history of a halo can be statistically analyzed using the Press-Schechter (PS) \citep{press1974} and the extended Press-Schechter (EPS) formalisms (see for example \citealt{vdb02,mcbride,correa1,correa2}).  A summary of the different merger tree-based methods for analyzing the evolution of dark matter halos is given in \citet{vdb14_1}. We evolve the halo mass according to the mass accretion history prescription as given by \citet{vdb14_2}. The virial mass $M_{200}$ as a function of redshift ($z$) for various present-day halo masses is shown in figure \ref{fig:halo_acc}.

The concentration parameter $c$ for the NFW profile is examined by \citet{zhao}, who found a significant dependence on the halo accretion and merger history. We use the expression given by \citet{vdb14_2} for the time evolution of $c$,
\begin{equation}
c=4 \left[ 1+ \left( \frac{t}{3.4 t_{0.04}} \right)^{6.5} \right]^{1/8} \; ,
\label{conc}
\end{equation}
where $t_{0.04}$ is the time at which the halo was $0.04$ times its present mass. 
\subsection{Gravitational potential}
The gravitational potential due to the dark matter distribution satisfies the Poisson equation
\begin{equation}
\label{poisson}
\nabla^2 \Phi_{\rm DM} =4\pi G (\rho-\rho_m),
\end{equation}
where $\rho=\rho(r)$ and $\rho_m$ is the mean matter density of the Universe. In general, it is difficult to construct an analytic expression for the potential corresponding to a given density, especially if the density is given by a complicated expression. However, since we can analytically differentiate even complicated expressions, fitting the potential and taking its derivatives to obtain the density (using equation \ref{poisson}) is a more convenient way to obtain a consistent analytic density-potential pair. 
\begin{figure*}
		\includegraphics[width=\columnwidth]{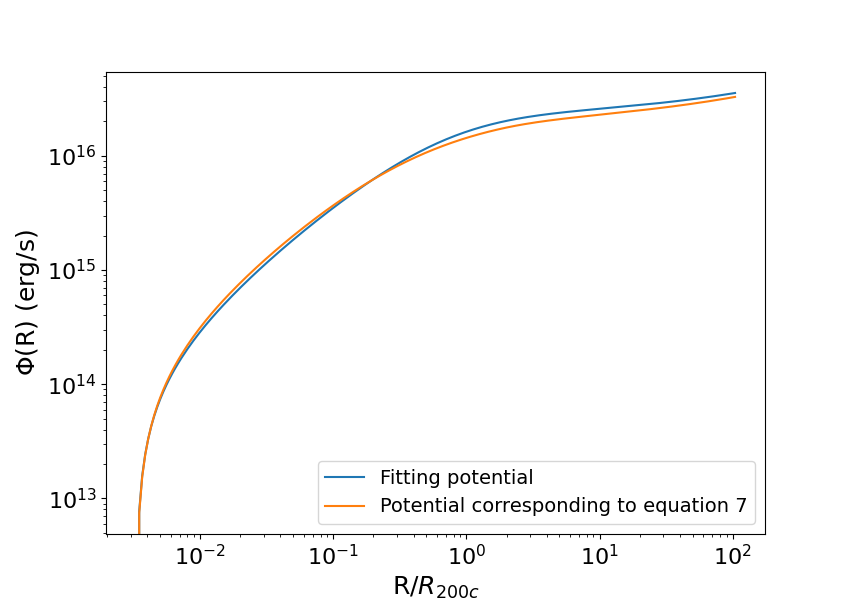}
	\hfill
		\includegraphics[width=\columnwidth]{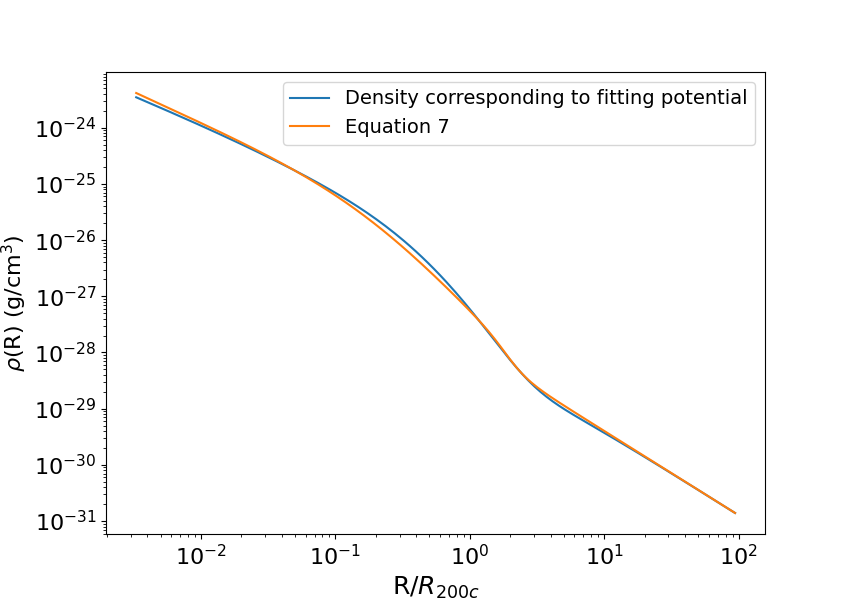}
	\caption{A comparison of the density-potential pair of the fitting potential (equation \ref{tot_potential}) and the true potential (numerical solution of equation \ref{poisson} with density given by equation \ref{eq:dk14last}), for a halo mass $M_0=10^{14} M_\odot$. Left panel shows the fitting potential (equation \ref{tot_potential}) and the true potential corresponding to equation \ref{eq:dk14last}. Right panel: density corresponding to fitting potential and true density as given by equation \ref{eq:dk14last}. Our choice of fitting potential results in a density-potential pair that is close (within $\sim$20\%) to the true values at all radii, and the corresponding acceleration ($- \partial \Phi/ \partial r$) has a maximum error of $\sim 30$ \%.}
	\label{fig:dens_pot}
\end{figure*}
We note that according to equation \ref{eq:dk14last}, $\rho(r) \rightarrow \rho_{\rm NFW}$ at small radii ($r << R_{200c}$) and $\rho(r) \rightarrow \rho_{\rm out}$ at large radii ($r >> R_{200c}$). Therefore, the corresponding gravitational potential can be well approximated by:
\begin{equation}
\Phi_{\rm DM,fit}=f_1 \Phi_{\rm NFW}(r)+f_2 \Phi_{\rm out}(r),
\label{tot_potential}
\end{equation}
where $\Phi_{\rm NFW}$ and $\Phi_{\rm out}$ correspond to the NFW and outer density potentials respectively, given by
\begin{equation}
\Phi_{\rm NFW}(r)=-\frac{4 \pi G \delta_c \rho_{\rm cr} R_s^3}{3} \frac{\ln(1+r/R_s)}{r} \; ,
\end{equation}
\begin{equation}
\Phi_{\rm out}(r)=4\pi G \rho_m \left( b_e \frac{r^{2-s_e}(5R_{200m})^{s_e}}{(3-s_e)(2-s_e)} +\frac{r^2}{6} \right) \; ,
\end{equation}
and $f_1$ and $f_2$ are weight functions that go to $0$ at large and small radii respectively. We substitute this in equation \ref{poisson} in order to obtain the density $\rho_{\rm fit}$ that corresponds to $\Phi_{\rm DM,fit}$. We try several functions $f_1$ and $f_2$ to match $\rho_{\rm fit}$ with $\rho(r)$ as given by equation \ref{eq:dk14last}. We find that the weight functions
\begin{equation}
    f_1= \frac{1-{\rm erf}(1.4x)}{2} \; ,\quad \quad f_2=1-f_1 \; ,
\end{equation}
where $x=\log_{10}(r/r_t)$, lead to a density-potential pair in reasonable agreement with equation \ref{eq:dk14last}. Figure \ref{fig:dens_pot} shows the values of $\Phi_{\rm DM,fit}$ and the corresponding $\rho_{\rm fit}$, as well as the "true" values as given by equations \ref{eq:dk14last} and the numerical solution of equation \ref{poisson}, for a halo mass $M_0=10^{14} M_\odot$. We note that the fitting potential closely matches the "true" potential at all radii. Therefore, we use equation \ref{eq:dk14last} to model the dark matter density distribution and equation \ref{tot_potential} to model its gravitational potential. We do not consider the gravity of the gas in our simulations. We note that if the gas has a similar profile as the dark matter the gravitational potential will also be similar, so the results will not be much different.

Cosmological simulations typically use comoving coordinates which automatically ensure that Hubble's law is satisfied. However, we use physical coordinates in our simulations. Therefore, we include a cosmological potential $\Phi_{\rm cos}$ to account for the Hubble expansion ($v=H[z] r$) far beyond the halo. In absence of halo gravity, the velocity of the uniform Universe should evolve as 
\begin{equation}
\frac{\partial v}{\partial t}=-v \frac{\partial v}{\partial r}-\frac{d\Phi_{\rm cos}}{dr} \; .
\end{equation}
For Hubble expansion, $-v \ \partial v/\partial r=-H^2 r$ and $\partial v / \partial t=\dot{H} r$. Therefore, for an evolution consistent with Hubble expansion, we have
\begin{equation}
g=(\dot{H}+H^2)r \; , \quad \quad \Phi_{\rm cos}=-(\dot{H}+H^2) \frac{r^2}{2} \; .
\end{equation}
For a flat Universe, this potential simplifies to 
\begin{equation}
\label{cosm_potential}
\Phi_{\rm cos}(r)=-H^2 \left( 1-\frac{3 \Omega_m(z)}{2}\right) \frac{r^2}{2} \; ,
\end{equation}
where $\Omega_m(z)=\Omega_m (1+z)^3$. Thus, the total potential is $\Phi(r)=\Phi_{\rm DM}(r)+\Phi_{\rm cos}(r)$.  Note that here $r$ is the physical (and not comoving) radial coordinate. With numerical simulations, we have verified that we do not obtain an evolution consistent with Hubble's law at large radii in absence of the cosmological potential in our model.

\subsection{Grid, initial and boundary conditions}
The baryonic gas is initialized at all radii as $\rho_{\rm g}=0.2 \ \rho_{\rm DM}$ at $z=6$ (such that $\rho_{\rm g}/[\rho_{\rm g}+\rho_{\rm DM}]=1/6$), and is evolved until $z=0$. The velocity is initialized as the Hubble velocity, i.e., $v(r)=H(z=6) \ r$. The temperature is initialized to a simple profile with high-temperature ($10^7 K$ for $M_0 \geq 10^{13} M_\odot$, the typical virial temperature of a cluster, and $10^6 K$ for $M_0<10^{13} M_\odot$, the typical virial temperature of the Milky Way halo) within the virial radius ($R_{200c}$) and a low temperature ($10^5 K$) outside it. Except for early transients and the value of the core entropy, our results are mostly independent of the initial temperature profile. We maintain a floor temperature of $2 \times 10^4$ K in our simulations, crudely mimicking UV heating of unshielded gas. 

We start with a series of 1D runs and then move on to 2D axisymmetric $(r,\theta)$ runs. We perform our simulations in spherical coordinates with $0 \leq \theta \leq \pi$ and $r_{\rm [min,max]}=[0.5,10^5]~\rm kpc$. The $\theta$ grid is uniformly spaced, and the radial grid is uniformly spaced in the innermost 10 kpc and logarithmically spaced outside this. The radial grid spacing in the inner regions of our simulations is 0.1 kpc, and the $\theta$ grid spacing is 3.6$^{\circ}$. This choice of radial grid gives us the necessary resolution to study both the inner and outer regions of the halo without very small code time steps (we have verified that we obtain similar volume-averaged quantities even at half of this resolution).

We apply outflow boundary conditions at the outer radial boundary. At the inner radial boundary, we impose $dP/dr=-\rho g$ (hydrostatic equilibrium) and restrict the gas from entering the computational domain (but it is allowed to leave). Reflective boundary conditions are applied in $\theta$. We note that cold gas has a tendency to artificially “stick” at the $\theta$ boundaries for our reflective boundary conditions. This cold gas can lead to an unphysically large accretion rate close to the poles, and hence
artificially enhanced feedback heating for runs with AGN feedback. This well-known effect was also encountered in \citetalias{prasad15}, and was quantified to some extent. We therefore exclude $0.15$ radians in the $\theta$ grid at each pole when estimating the mass accretion rate used to calculate the power injection by the jets in the feedback runs. 
\subsection{Radiative cooling} 
\label{sec:cooling}
The cooling rate of gas and plasmas is typically dominated by collisional processes, and can be expressed as 
\begin{equation}
\mathcal{L}=n_e n_i \Lambda(T,Z)
\end{equation} 
where $n_e$ and $n_i$ and the electron and ion densities respectively, and $\Lambda(T,Z)$ is the cooling function. The value of $\Lambda$ depends on several factors, including temperature, metallicity of the gas, number density of electrons, and the presence of a radiation field \citep{wiersma2009,schure,wang}. 

In our runs, we use the cooling function of \citet{wiersma2009}, which was calculated by running photoionization models and calculating the element-by-element cooling rates under exposure to the cosmic microwave and UV/X-ray radiation. The total cooling function can be expressed as a sum of the contributions from Hydrogen, Helium, and heavier elements (equation 4 in \citealt{wiersma2009}),
\begin{equation}
\label{cooling_eqt}
\Lambda=\Lambda_{\rm H,He}+\Sigma_{i> \rm He} \Lambda_{i,\odot} \left( \frac{n_e}{n_{e,\odot}} \right) \left(\frac{n_i}{n_{i,\odot}} \right) \; .
\end{equation}
The cooling function strongly depends on the metallicity of the gas. The intergalactic medium (IGM) is polluted by metals produced in massive stars via their winds and after their death as supernovae (e.g., \citealt{nath,wiersma}). These metals are transferred out via galactic outflows and mixed further due to mergers. The chemical enrichment of the IGM is modelled by hydrodynamical simulations of outflows from star-forming galaxies \citep{oppenheimer}, and further turbulent diffusion of metals \citep{shen}. We use a simple analytic fit for the metallicity as a function of redshift that corresponds to figure 8 of \citet{wiersma} and crudely mimics the metallicity within and outside halos:
\begin{equation}\label{metal}
\log_{10} \left( \frac{Z(z)}{Z_\odot} \right)=
\begin{cases}
-0.52-0.25 z \quad \quad r<R_{200} \\
-2-0.5 z \quad \quad r>R_{200} \; .
\end{cases}
\end{equation}
This fit evolves the metallicity within the halo from $0.01 Z_\odot$ at $z=6$ to $0.3 Z_\odot$ at $z=0$. The value of $0.3 Z_\odot$ at the present redshift is chosen to match observations of the cluster outskirts/IGM. The metallicity is kept low outside the halo, evolving from $10^{-5} Z_\odot$ at $z=6$ to $10^{-2} Z_\odot$ at $z=0$.
\subsection{Evolution of the central supermassive black hole}
\label{sec:smbh}
We place a seed black hole at the center of the halo at $z=6$ and allow it to grow due to accretion of the infalling gas. The mass inflow rate $\dot{M}_{\rm in}$ is calculated as the cold gas mass accretion rate at a radius of $\approx 1$ kpc:
\begin{equation}
    \dot{M}_{\rm in}=r_{\rm in}^2 \int \rho v d\Omega \quad \text{for cold gas } (T<10^5 K) \; ,
\end{equation}
where $d \Omega=2\pi \sin \theta d \theta$. 
Observations show that only a tiny fraction of the mass accreting at $\sim$ 1 kpc makes it all the way down to the central SMBH (the best evidence for this is the Galactic center black hole; e.g., see  \citealt{baganoff2003}). Since our simulations do not resolve sub-kpc scales, we assume a small fraction of $\dot{M}_{\rm in}$ accretes on to the SMBH
\begin{equation}
\label{eq:mdot}
\dot{M}_{\rm acc}=\epsilon_m \dot{M}_{\rm in} \; ,
\end{equation}
where $\epsilon_m$ is  the efficiency of mass transport from the inner boundary at $\sim$ 1 kpc to the BH event horizon. We limit $\dot{M}_{\rm acc}$ by the Eddington limit of the central SMBH, given by 
\begin{equation}
\label{eq:MdotEdd}
\dot{M}_{\rm Edd}=\frac{4\pi G M_{\rm BH} m_p}{\epsilon_{\rm BH} \sigma_T c} \; ,
\end{equation}
where $\epsilon_{\rm BH}$ is the black hole mass-to-energy conversion efficiency, taken to be the fiducial value of 0.1. The accretion rate onto that BH is then taken as the minimum of that calculated in equations \ref{eq:mdot} and \ref{eq:MdotEdd}. The mass transport efficiency $\epsilon_m$ is kept constant during a run, and for simplicity we assume  $\epsilon_{\rm BH} \dot{M}_{\rm acc} c^2$ to be the feedback power. The evolution of mechanical and radiative output of a BH for different accretion rates is studied in \citealt{churazov}. 

In each computational step, the mass of the black hole grows by 
\begin{equation}
\label{eq:dM}
\Delta M_{\rm BH}=\dot{M}_{\rm acc} \times \Delta t \; .
\end{equation}
Thus, the BH grows faster during core cooling (hence a large $\dot{M}_{\rm in}$).  
We note that massive (cluster-sized) halos contain a large number of massive galaxies, each with their own SMBHs. In our massive halo simulations, we only consider the feedback due to the central SMBH of the brighest central galaxy (BCG).  
\subsection{AGN jet feedback}
\label{sec:agn_implementation}
We implement AGN feedback in a manner similar to \citetalias{prasad15}. Here we summarize the important features. The density source term (see equation \ref{eqt_density}) is given by 
\begin{equation}
\label{source_dens}
    S_\rho(r,\theta)=\mathcal{N} \dot{M}_{\rm jet} \psi(r,\theta) \; ,
\end{equation}
where $\dot{M}_{\rm jet}$ is the \textit{single-jet} mass loading rate, $\psi$ is a geometrical factor that smoothly falls to zero outside the biconical region of radius $r_{\rm jet}$ and half-opening angle $\theta_{\rm jet}$ and $\mathcal{N}$ is a normalization factor. The exact expressions for $\psi(r,\theta)$ and $\mathcal{N}$ are given in equation 5 of \citetalias{prasad15}. We select the parameters $r_{\rm jet}= 2$ kpc, $\theta_{\rm jet}=\pi/6$, $\sigma_r=0.05$ kpc, $\sigma_\theta=0.05$, and $v_{\rm jet}=3 \times 10^4$ km s$^{-1}$ ($0.1c$) for all our simulations. The results are largely insensitive to the exact values of these parameters. The mechanical efficiency of feedback and mass transfer efficiency, however, are important parameters that strongly influence the evolution. 

The total energy released by jets is
\begin{equation}
    \dot{E}_{\rm jet}=\dot{M}_{\rm jet} v_{\rm jet}^2 =  \epsilon_{\rm BH}\dot{M}_{\rm acc} c^2 \; .
\end{equation}
Out of this, a fraction $f_{\rm kin}=0.5$ is input as kinetic energy and the remaining is input as thermal energy. The total feedback efficiency relating the feedback mechanical power and the accretion rate at $\sim 1$ kpc ($\dot{E}_{\rm jet} = \epsilon \dot{M}_{\rm in} c^2$) is therefore 
\begin{equation}
    \epsilon=\epsilon_m \times \epsilon_{\rm BH}.
\end{equation}
Our AGN feedback model with a small $\epsilon$ can also crudely mimic stellar feedback in the central galaxy, which has a lower efficiency with respect to the mass cooling rate (e.g., see figure 4 and its discussion in \citealt{sharma12b}). Stellar feedback, and not AGN feedback, is expected to regulate star formation in lower mass halos ($\lesssim 10^{12} M_\odot$).

\section{Non-radiative runs} \label{sec:nonrad}
\begin{figure*}
	\includegraphics[width=2\columnwidth]{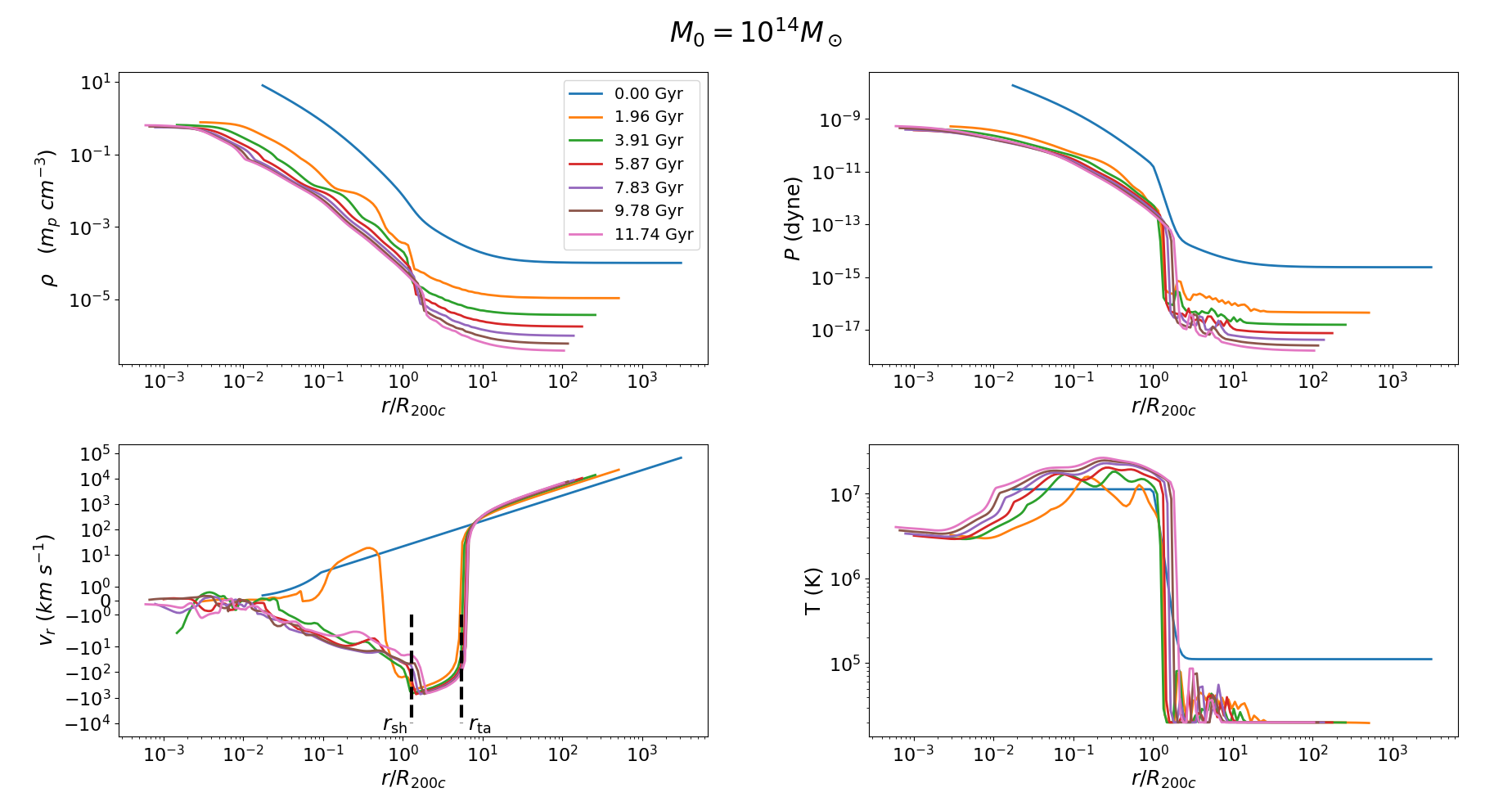}
	\caption{Density, pressure, velocity, and temperature profiles at various times for the non-radiative runs with $M_0=10^{14} M_\odot$. All the panels show the profiles at the times indicated in the legend in the density panel. Note that the x-axis in each plot is normalized to $R_{200c}$ at that time. The inner halo ($<0.01 R_{200c}$) gas shows a core, with a nearly constant temperature and density. The shock radius ($r_{\rm sh}$) and the turnaround radius ($r_{\rm ta}$) are shown in the velocity profile plot. The outer radii ($>10 R_{200c}$) are driven by Hubble expansion, with velocity linearly increasing with radius as $v=H(z)r$. This adiabatic expansion leads to a low temperature in the outer regions, where we impose a floor of $2 \times 10^4$ K corresponding to the IGM temperature after reionization.}
	\label{fig:basic}
\end{figure*}
We start with 1D simulations in the absence of radiative cooling and AGN feedback that serve as control runs for us to interpret more complex simulations presented later. Figure \ref{fig:basic} shows the density, pressure, velocity, and temperature profiles at different times for a halo mass $M_0=10^{14} M_\odot$. In the absence of heating and cooling, a core is formed that extends up to $\sim 10^{-2} R_{200c}$, similar to the results of \citet{choudhury}. The core has a lower temperature (few $\times 10^6 K$ for $M_0=10^{14} M_\odot$) compared to the virial radius. It is also characterized by a flatter density profile.

In the absence of cooling, the velocity within the virial radius is subsonic and its magnitude decreases with decreasing radius.  
The gas behaves adiabatically within the shock radius, where the gravity due to the dark matter halo is much stronger than the gravity due to the cosmological potential (equation \ref{cosm_potential}). At all times, $\dot{M}_{\rm in}=0$ because of the lack of gas with $T<10^{5} K$ (moreover, the gas is almost hydrostatic). Therefore, there is no growth of the central SMBH according to our BH growth prescription (see equations \ref{eq:mdot}, \ref{eq:dM}). 

\textit{Hubble Expansion:} Beyond $10 R_{200}$, the velocity is completely radially outward and follows Hubble expansion, $v(r)=H(z) r$. This is because of the cosmological potential (equation \ref{cosm_potential}) which ensures consistency with Hubble's law on large scales. As $H(z)= H_0 E(z)$ (where $E(z)=\sqrt{\Omega_m (1+z)^3 + \Omega_\Lambda}$), the velocity at the outermost point of the computational domain evolves as $v_{\rm out} \propto r_{\rm out} E(z)$.

\textit{Formation of the Virial Shock:} The virial shock forms through multiple stages (see \citealt{dekel}). Because of our choice of the initial temperature/entropy profiles, shocks are formed at early times but they are not stationary and deviate from similarity solutions. 
After $z \approx 3.1$ a virial shock is fully localized near $R_{200c}$. This shock formation redshift is smaller for more massive halos because of their deeper potential wells. The gas in massive halos therefore falls at supersonic speeds ($v>c_s$), forming a stronger shock compared to less massive halos. 
\subsection{Baryon fraction evolution} \label{bf_evolution}
Our simulations do not model star formation, and all of the baryonic matter is treated as gas. Therefore, we approximate the baryon fraction within a radius $r$ as the gas fraction within that radius, namely
\begin{equation}
f_b=\frac{M_{\rm g}(<r)}{M_{\rm g}(<r)+M_{\rm DM}(<r)} \; ,
\end{equation}
with $M_{\rm g}$ and $M_{\rm DM}$ being the gas and dark matter mass within the radius $r$, respectively. Recall that our initialization condition is $f_b(z=6)=1/6$ (the universal baryon fraction) at all radii.

\begin{table}
	\begin{center}
		\caption{Outer density parameters (see equations \ref{eq:dk14first}-\ref{eq:dk14last}) for maintaining universal baryon fraction}
		\label{tab:dk14_parameters}
		\begin{tabular}{|c|c|c|} 
			\hline
			Halo Mass at $z=0$ ($M_0$) & $s_e(z)$ & $b_e(z)$ \\
			\hline
			$10^{12} M_\odot$ & $1.4-0.033 z$ & $1.2+0.083 z$ \\
			\hline
			$10^{13} M_\odot$ & $1.6-0.05 z$ & $1.1+0.417 z$ \\
			\hline
			$10^{14} M_\odot$ & $1.5-0.033 z$ & $1.2+0.583 z$ \\
			\hline
			$10^{15} M_\odot$ & $1.1-0.016 z$ & $2.0+0.5 z$ \\
			\hline
		\end{tabular}
	\end{center}
\end{table}

The baryon fraction at the virial radius depends on the infall of gas across $R_{200c}$, which in turn depends on the gravitational field due to the dark matter halo. For a density varying as $\rho(r) \propto r^\alpha$, the gravitational acceleration follows $g(r) \propto r^{\alpha+1}$. In particular, we find that the parameters $b_e$ and $s_e$ characterizing the outer density profile (see equation \ref{eq:dk14_out}) strongly affect the baryon fraction near the virial radius.

We test various values of $b_e$ and $s_e$ for different halo masses, and find that static values of these parameters ($b_e=2.5$, $s_e=1.5$) lead to a baryon fraction well below the universal value. However, we show that we can closely match the universal baryon fraction within $R_{200c}$ at all times by introducing redshift dependence in the parameters. In table \ref{tab:dk14_parameters}, we list the expressions for $s_e$ and $b_e$ that were found to achieve close to the universal baryon fraction near and outside $R_{200c}$. These parameters increase the DM density outside the halo and increase the gravitational pull of the dark matter to enhance the gas mass accretion rate across $R_{200c}$.

\begin{figure}
	\includegraphics[width=\columnwidth]{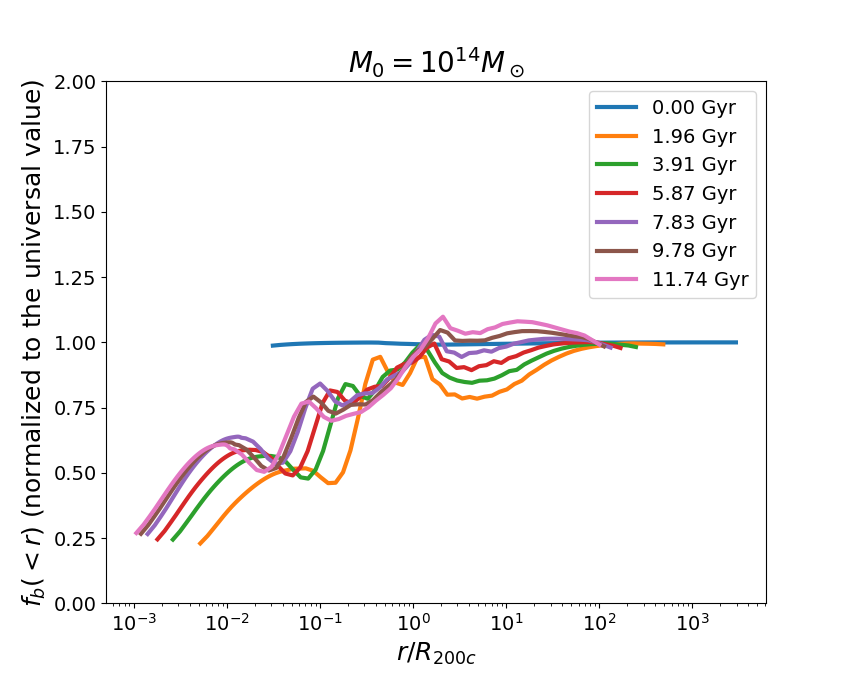}
	\caption{Baryon fraction evolution for a non-radiative run with $M_0=10^{14} M_\odot$, using the redshift-dependent parameters for $s_e$ and $b_e$ given in  table \ref{tab:dk14_parameters}. Using these values, the baryon fraction at the virial radius is close to the universal value at all times. As the parameters $s_e$ and $b_e$ characterize the outer DM density profile (see equation \ref{eq:dk14_out}), the baryon fraction at the inner radii ($r<10^{-1} R_{200c}$) is largely independent of the gravity in the outer regions and depends more on the radiative/feedback effects discussed later.}
	\label{fig:new_bf}
\end{figure}

In figure \ref{fig:new_bf}, we plot the baryon fraction as a function of radius at different times for our non-radiative/non-feedback runs for a halo mass $M_0=10^{14} M_\odot$, using redshift-dependent parameters for $b_e$ and $s_e$ listed in table \ref{tab:dk14_parameters}. We see that the baryon fraction lies within $\sim 20\%$ of the universal value for $r \gtrsim R_{200c}$ at all times $t \gtrsim 2$ Gyr ($z \lesssim 2$). 
\subsection{Self-Similar evolution}
\begin{figure}
	\includegraphics[width=\columnwidth]{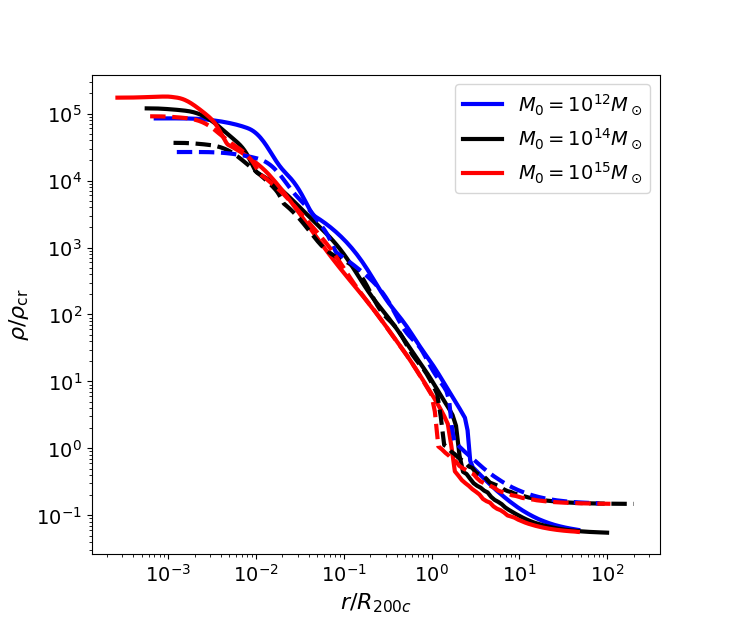}
	\caption{Scaled density ($\rho/\rho_{\rm cr}$) versus scaled radius ($r/R_{200c}$) for different halo masses at $z=1$ (dashed lines) and $z=0$ (solid lines). In the absence of radiative cooling and feedback, the density profiles outside the core are reasonably self-similar at all times after the initial transients. The inner profiles are self-similar in $r/R_{200c}$ and the outer ones in $r/R_{200m}$.}
	\label{fig:self_similar}
\end{figure}
Figure \ref{fig:self_similar} shows the gas density as a function of radius for various halo masses at a redshift of $z=1$ ($t=5.06$ Gyr) and $z=0$ ($t=12.90$ Gyr). The density is scaled with respect to the critical density of the universe $\rho_{\rm cr}$, and the radius is scaled with respect to $R_{200c}$ of each halo at that redshift. The inner regions show variations in the profile due to the differences in the core entropy. However, we can see that the regions $>0.03 R_{200c}$ behave reasonably self-similarly with respect to $R_{200c}$.\\
This self-similar behavior of the gas across a large range of halo masses occurs because gravitational field due to the inner (NFW) dark matter profile (equation \ref{eq:dk14first}) and the outer profile (equation \ref{eq:dk14_out}) are both self-similar with respect to $R_{200c}$ and $R_{200m}$ respectively. We therefore expect the gas to behave self-similarly outside the core (after the initial transients). The deviations from self-similarity are the greatest near the virial shock ($R_{200c}$), where the inner and outer DM density profiles are patched together. The concentration parameter $c$ and the formation redshift (equation \ref{conc}) depend on the halo mass, so self-similarity is not exact. We note that our density profile goes to $\rho_m$ and not $\rho_{\rm cr}$ at large radii (a consequence of equation \ref{eq:dk14_out}), so self-similarity beyond $R_{200c}$ is not present. However, we note that if we scale the density with respect to $\rho_m$ and the radius with respect to $R_{200m}$, the outer regions beyond the halo are self-similar, while the inner regions are not. This is in agreement with \citetalias{dk14}. 

\begin{table*}
    \begin{center}
    \begin{tabular}{c|c|c|c|c|c|c}
         \hline
         \textbf{Type} & $M_0$ ($M_\odot$) & $R_{200c}$ (kpc) & $\epsilon_m$ & avg($\dot{M}_{\rm in}/\dot{M}_{\rm in,CF})$ & $f_b(R_{200c},\rm{z=0})$ & $M_{\rm BH,z=0}$ ($M_\odot$)\\
         \hline
         NR & $10^{14}$ & 959.09 & $10^{-3}$ & 0 & 0.158 & $10^6$ \\
         \hline
         CF & $10^{14}$ & 959.09 & $10^{-3}$ & 1 & 0.142 & $2.63 \times 10^{9}$ \\
         \hline
         AGN & $10^{14}$ & 959.09 & $10^{-5}$ & 0.969 & 0.139 & $3.38 \times 10^7$ \\
         \hline
         AGN & $10^{14}$ & 959.09 & $10^{-4}$ & 0.545 & 0.138 & $1.87 \times 10^8$ \\
         \hline
         AGN & $10^{14}$ & 959.09 & $10^{-3}$ & 0.160 & 0.162 & $1.05 \times 10^9$ \\
         \hline
         NR & $10^{12}$ & 206.62 & $10^{-3}$ & 0 & 0.137 & $10^6$ \\
         \hline
         CF & $10^{12}$ & 206.62 & $10^{-3}$ & 1 & 0.032 & $3.46 \times 10^7$ \\ 
         \hline
         AGN & $10^{12}$ & 206.62 & $10^{-6}$ & 0.853 & 0.012 & $1.19 \times 10^6$ \\
         \hline
         AGN & $10^{12}$ & 206.62 & $10^{-5}$ & 0.353 & 0.059 & $1.86 \times 10^6$ \\
         \hline
         AGN & $10^{12}$ & 206.62 & $10^{-4}$ & 0.167 & 0.043 & $8.03 \times 10^6$ \\
         \hline
         AGN & $10^{12}$ & 206.62 & $10^{-3}$ & 0.040 & 0.027 & $8.22 \times 10^6$ \\
         \hline
    \end{tabular}
    \caption{Results from some of our simulations for non-radiative (NR), cooling flow (CF), and AGN feedback (AGN) runs. In the runs tabulated above, the seed black hole mass is kept fixed at $10^6 M_\odot$, and the present-day halo mass $M_0$ and mass transport efficiency $\epsilon_m$ are varied. It is clear that the average cold gas mass inflow rate ($\dot{M}_{\rm in}$) decreases with increase in feedback efficiency. The baryon fraction at $R_{200c}$ varies only slightly for the $10^{14}$ halo, while it strongly depends on whether radiative cooling and feedback are included in the $10^{12}$ halo. The mass of the central SMBH at $z=0$ is the highest in the cooling flow case. For the feedback runs, the SMBH mass increases with increase in $\epsilon_m$ (see the discussion in section \ref{var_eff}).}
    \label{tab:runs_table}
    \end{center}
\end{table*}
\section{Runs with radiative cooling} 
\label{sec:cooling_runs}
We now present results from our runs with radiative cooling (but no feedback heating). The metallicity dependent cooling function is described in section \ref{sec:cooling}, and we evolve the metallicity of the computational domain (halo+intergalactic medium) according to equation \ref{metal}. We use the initial conditions  described in section \ref{sec:nonrad}, namely,  $\rho_{\rm g}=0.2 \rho_{\rm DM}$ at $t=0$ ($z=6$). 
\subsection{Lower mass halos}
\begin{figure*}
	\includegraphics[width=2\columnwidth]{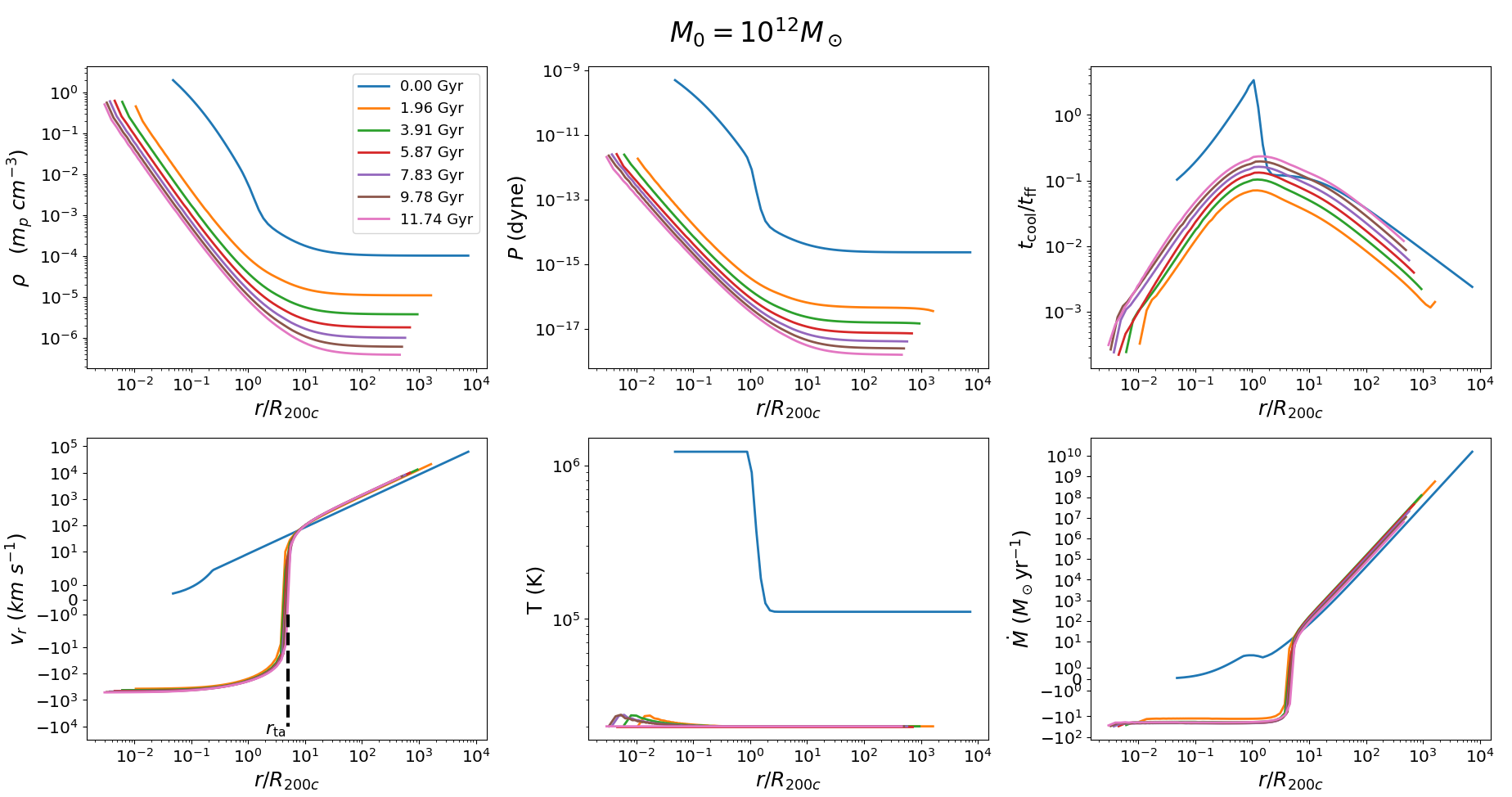}
	\caption{Cooling flow profiles for $M_0=10^{12} M_\odot$, showing the same quantities as figure \ref{fig:basic}, with the addition of $t_{\rm cool}/t_{\rm ff}$ (top right) and mass accretion rate $\dot{M}$ (bottom right) profiles versus scaled radius $r/R_{200c}$. From the density and temperature profiles, it is evident that the gas quickly cools to the floor temperature and no virial shock is formed. The velocity of the gas within the turnaround radius ($r_{\rm ta}$) is completely inward, giving rise to a large inflow rate at the center of the halo. This low mass halo has a $t_{\rm cool}/t_{\rm ff} \lesssim 0.1$ within the halo, implying that cooling dominates within $R_{200c}$. Note that $\Lambda$ is discontinuous at $2 \times 10^4$ K because of our temperature floor, so although $t_{\rm cool}/t_{\rm ff}<0.1$, its exact value can be different from what is indicated.}
	\label{fig:12cooling}
\end{figure*}

Figure \ref{fig:12cooling} shows various profiles for a $M_0=10^{12} M_\odot$ halo with radiative cooling. In addition to the profiles shown in figure \ref{fig:basic}, we also show the $t_{\rm cool}/t_{\rm ff}$ and mass accretion rate profiles. With radiative cooling, lower mass halos are unable to form a stable pressure-supported virial shock. After the initial transient phase, cooling dominates at all radii $<10 R_{200c}$, and the temperature drops to the floor temperature at all radii. 

The velocity is radially inwards everywhere within  $10 R_{200c}$ (the turn-around radius), outside which Hubble expansion and the associated adiabatic cooling dominate. There is no change in velocity near $R_{200c}$ typically associated with a virial shock. The mass accretion rate is fairy constant across radii within the turnaround radius (bottom right panel of figure \ref{fig:12cooling}). 

We can understand the behavior of the cooling flow runs by analysing some important timescales. 
These timescales are given by 
\begin{equation}
t_{\rm cool}=\frac{3}{2}\frac{ n k_B T}{n_i n_e \Lambda(T)} \; ,
\end{equation}
\begin{equation}
t_{\rm ff}=\left( \frac{2r}{g}\right)^{1/2} \; .
\end{equation}
The gas in the outer regions far beyond the halo is dominated by the cosmological potential, which gives an expansion timescale of 
\begin{equation}
    t_{\rm exp}=\frac{1}{H}.
\end{equation}
The top right panel of figure \ref{fig:12cooling} plots $t_{\rm cool} / t_{\rm ff}$ as a function of radius at different times. The ratio $t_{\rm cool}/t_{\rm ff} \ll 1$ at all radii and times, indicating that radiative cooling dominates at all radii (also seen in the temperature profile which is isothermal at the floor temperature). 
The roughly constant values $\dot{M}$ (a coincidence since the cosmological decrease in density is compensated by an increase in $R_{200c}$), temperature and infall velocity imply that the density and pressure decrease as $r^{-2}$. For $\rho \propto r^{-2}$, the infall time ($t_{\rm in} = r/|v_r|$) is comparable to the cooling time (see equation A3 in \citealt{sharma12}), but shorter than $t_{\rm ff}$ (for a similar situation, see $r <2$ kpc in figure A1 of \citealt{sharma12}). Equation A2 in \citet{sharma12} implies that $|d\ln v/d\ln r| \ll 1$ for $t_{\rm cool}/t_{\rm ff} \ll 1$, consistent with the bottom-left panel of figure \ref{fig:12cooling}.

The ratio $t_{\rm cool}/t_{\rm ff} \ll 1$ and a temperature much smaller than the virial temperature imply that the gas falls in at the free-fall speed. Far beyond the halo, the cosmological potential dominates halo gravity ($t_{\rm exp} << t_{\rm ff}$), so the velocity is consistent with Hubble expansion.

\subsection{Massive halos}
\begin{figure*}
	\includegraphics[width=2\columnwidth]{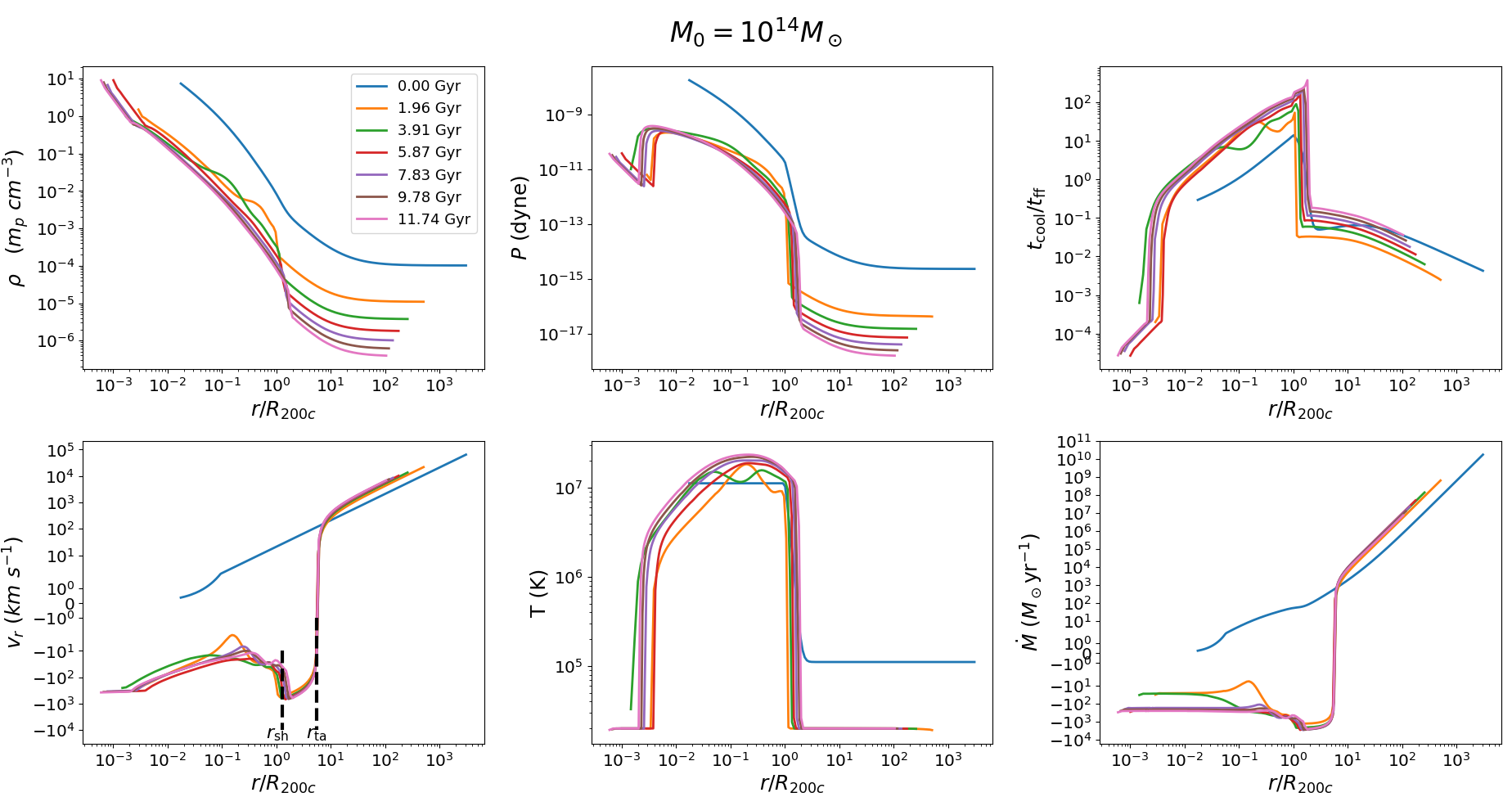}
	\caption{Cooling flow profiles for $M_0=10^{14} M_\odot$. 
	For this halo mass, radiative cooling strongly affects only the inner regions, where the density profile is cuspy instead of core-like and the temperature is much smaller compared to the non-radiative runs (bottom right panel of figure \ref{fig:basic}). However, the regions $>0.1 R_{200c}$ behave very similar to the non-radiative runs, and a stable virial shock is still formed close to $R_{200c}$.}
	\label{fig:14cooling}
\end{figure*}

Figure \ref{fig:14cooling} shows the profiles for the cooling flow run of a $M_0=10^{14} M_\odot$ halo. For higher mass halos ($M_0 \gtrsim 10^{13} M_\odot$), radiative cooling dominates only in the inner regions of the halo ($r < 0.01 R_{200}$), resulting in a core temperature significantly lower than that of the corresponding non-radiative run (see \ref{fig:basic}). The matter near the center is falling freely, and the density shows a cusp-like behavior instead of a core. 

The top right panel of figure \ref{fig:14cooling} shows the $t_{\rm cool} / t_{\rm ff}$ profile for this run, which suffers catastrophic cooling with $t_{\rm cool}/t_{\rm ff}<1$ only within $r<0.01 R_{200c}$.
However, its value for $r \gtrsim 0.1 R_{200c}$ is $\gg 1$, implying the formation of a stable virial shock. A growing subsonic, roughly hydrostatic, cooling flow develops outside the cuspy center. A steady cooling flow characterized by a roughly constant $\dot{M}$ (see the bottom-right panel) develops within a radius where the cooling time equals the age of the Universe.

By analyzing the stability of the virial shock, \citet{dekel} concluded that shocks do not form in smaller halos ($\lesssim$ few $\times 10^{11} M_\odot$). This transition mass depends on the metallicity of the gas; for $Z\sim 0.05 Z_\odot$ they found that this lower bound rises to $\sim 7 \times 10^{11} M_\odot$. We find a similar (but slightly higher) limit; there is no virial shock for $M_0=10^{12} M_\odot$. This small difference may be due to an increasing IGM/CGM metallicity with time in our simulations, which increases the cooling strength. 

\citet{stern2019} classify their cooling flows according to $R_{\rm sonic}$, the radius where the velocity of the gas equals the speed of sound in the medium, i.e. $\mathcal{M}=1$. For $M_0=10^{12}$, $R_{\rm sonic}$ is in the $\sim 100-1000$ kpc range in our runs (outside $R_{200c}$), which is higher than their choice of a cooling flow in galaxy-sized halos. They note that if $R_{\rm sonic}>R_{1/2}$ (the maximum radius at which the halo gas can be supported by gravity against angular momentum), then all the gas quickly collapses on a dynamical timescale, which is what happens in our runs. 

For a massive cluster $M_0=10^{15} M_\odot$, $R_{\rm sonic}$ ranges from $\sim 0.5$ kpc to $\sim 2.5$ kpc in our runs,\footnote{The existence of sonic point also depends on the nature of gravitational potential; e.g., the addition of an isothermal potential in the center of a cluster reduces $t_{\rm ff}$ and pushes the sonic radius further in (see figure 3 in \citealt{prasad2020}).} which is consistent with \citet{stern2019}. We show the profiles for $M_0=10^{14} M_\odot$ in figure \ref{fig:14cooling}, but the profiles for $M_0=10^{15} M_\odot$ are very similar. The flow is subsonic within the virial shock (except in the innermost few kpc, where cooling is very strong). Outside this, the infall time roughly equals the cooling time, which is longer than the free-fall time. The density power law index in the range $3 \times 10^{-3} R_{200c}-10^{-1} R_{200c}$ varies from $-1.5$ to $-1.25$ in our runs, in agreement with their result of $n_H \propto r^{-1.4}$. 

\section{Runs with AGN feedback} \label{sec:AGN}
In this section, we introduce AGN jet feedback in our runs with radiative cooling. The outer density for the DM halo still evolves according to table \ref{tab:dk14_parameters} (see equation \ref{eq:dk14_out}) so that the baryon fraction within the virial radius is close to that universal value. Table \ref{tab:runs_table} shows the list of important quantities for the feedback runs, as well for the corresponding cooling flow and non-radiative runs. Unlike earlier idealized simulations (e.g. \citealt{gaspari2012,li2014a}; \citetalias{prasad15}; \citealt{Li15}), the mass of the DM halo, central black hole mass, and other cosmological quantities are evolved in time. 
\begin{figure*} 
	\includegraphics[width=2\columnwidth]{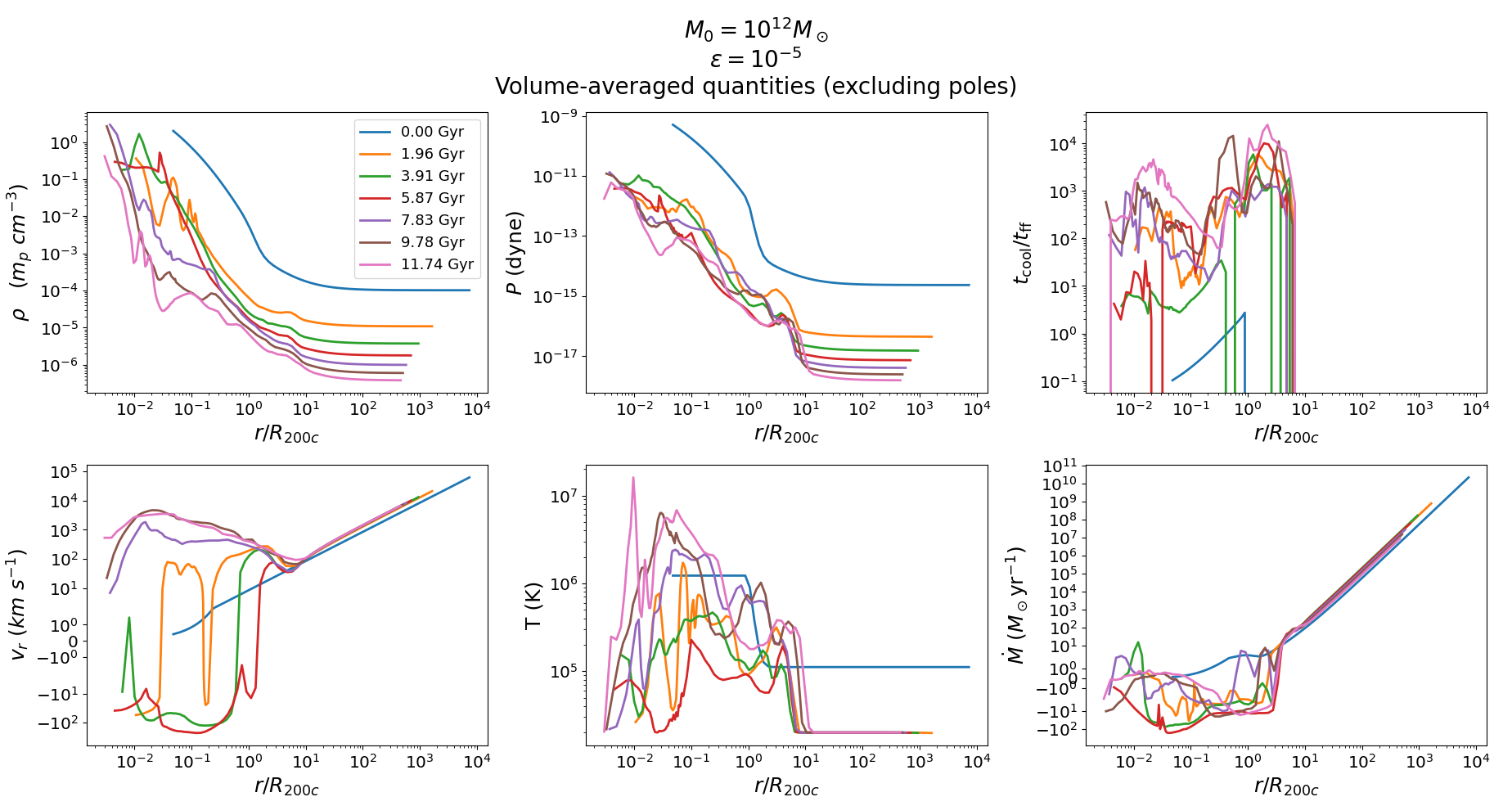}
	\caption{Flow profiles for the AGN feedback run for $M_0=10^{12} M_\odot$ and $\epsilon=10^{-5}$. We exclude $0.15$ rad at each pole in the $\theta$ direction while calculating the averages, because the gas has a tendency to artificially "stick" to the poles. Feedback heating prevents catastrophic cooling and temperature is not at the floor value. Since there is very little gas in the range 0.5-8 keV, we use the temperature range 0.1-8 keV to calculate the volume-weighted $t_{\rm cool}/t_{\rm ff}$. A pressure-supported shock is able to form due to feedback heating, but is pushed outward to $\sim 7 R_{200c}$. The mass inflow rate is smaller than the cooling flow run (see figure \ref{fig:12cooling}).}
	\label{fig:12_agn_profiles}
\end{figure*}

We select the run with $M_0=10^{14} M_\odot$, feedback efficiency $\epsilon=10^{-4}$ ($\epsilon_{\rm BH}=10^{-1}$ and $\epsilon_m=10^{-3}$), and seed black hole mass $M_{\rm BH,0}=10^6 M_\odot$ as our fiducial AGN feedback run. This results in a mass inflow rate of $\sim 16 \%$ compared to the corresponding cooling flow run, which is in broad agreement with observations of strong cool core clusters (e.g., see figure 3 in \citealt{mcdonald18}). 

The various quantities of interest to us are the cold gas mass within 5 kpc, the black hole mass accretion rate as defined in equation \ref{eq:mdot}, the jet power, and the minimum value of the volume-weighted $t_{\rm cool}/t_{\rm ff}$. The value of ${\rm min}(t_{\rm cool}/t_{\rm ff})$ is calculated similar to \citetalias{prasad15}. We only include the hot gas in our calculations (0.5 keV - 8 keV), which corresponds to the temperature range of X-ray emitting plasma. The jet power is also estimated in a manner similar to \citetalias{prasad15}. We consider the grid points with a temperature greater than the threshold value of $5\times 10^7$ K to belong to the bubble/jet material. We then volume-integrate the internal energy density of all cells belonging to the jet. The total internal energy is then divided by the lifetime of the AGN bubble, taken to be 30 Myr, to obtain the jet power estimate. 

\textit{Low mass halos:}
Figure \ref{fig:12_agn_profiles} shows the average density, pressure, $t_{\rm cool}/t_{\rm ff}$, velocity, temperature, and mass accretion rate profiles for a $M_0=10^{12} M_\odot$ run with feedback efficiency $\epsilon=10^{-5}$. For a pure cooling flow in a $M_0=10^{12} M_\odot$ halo, the gas quickly cools to the floor temperature and no virial shock is formed (figure \ref{fig:12cooling}). AGN feedback heating prevents catastrophic cooling, and supports the formation of a shock even in low mass halos. However, the shock is no longer near $R_{200c}$ and is pushed further outwards to $\sim 7 \times R_{200c}$; such a proposal can be tested via thermal Sunyaev-Zeldovich effect (e.g., \citealt{bregman21}) or the Fast Radio Bursts (FRBs; \citealt{macquart20}). Unlike the virial shock in massive halos, the location of the feedback-driven shock is expected to depend on the feedback efficiency. We note that the temperature no longer drops to the floor temperature within the halo, and a hot CGM primarily sustained by feedback heating is formed (e.g., see \citealt{sokolowska18}). In fact, the temperature at some points during the heating phase (characterized by a positive radial velocity) is even higher than the corresponding virial temperature of $\sim 10^6$ K. However, there is very little gas in the 0.5-8 keV range. We therefore use a temperature range of 0.1-8 keV to calculate the volume-weighted $t_{\rm cool}/t_{\rm ff}$, which shows large fluctuations in the core due to multiphase gas, but reaches larger and more uniform value near $R_{200c}$. A feedback efficiency $\epsilon \sim 10^{-7}$ is very similar to a pure cooling flow and $\epsilon \sim 10^{-6}$ shows a shock close to $R_{200c}$. 

\begin{figure*}
	\includegraphics[width=2\columnwidth]{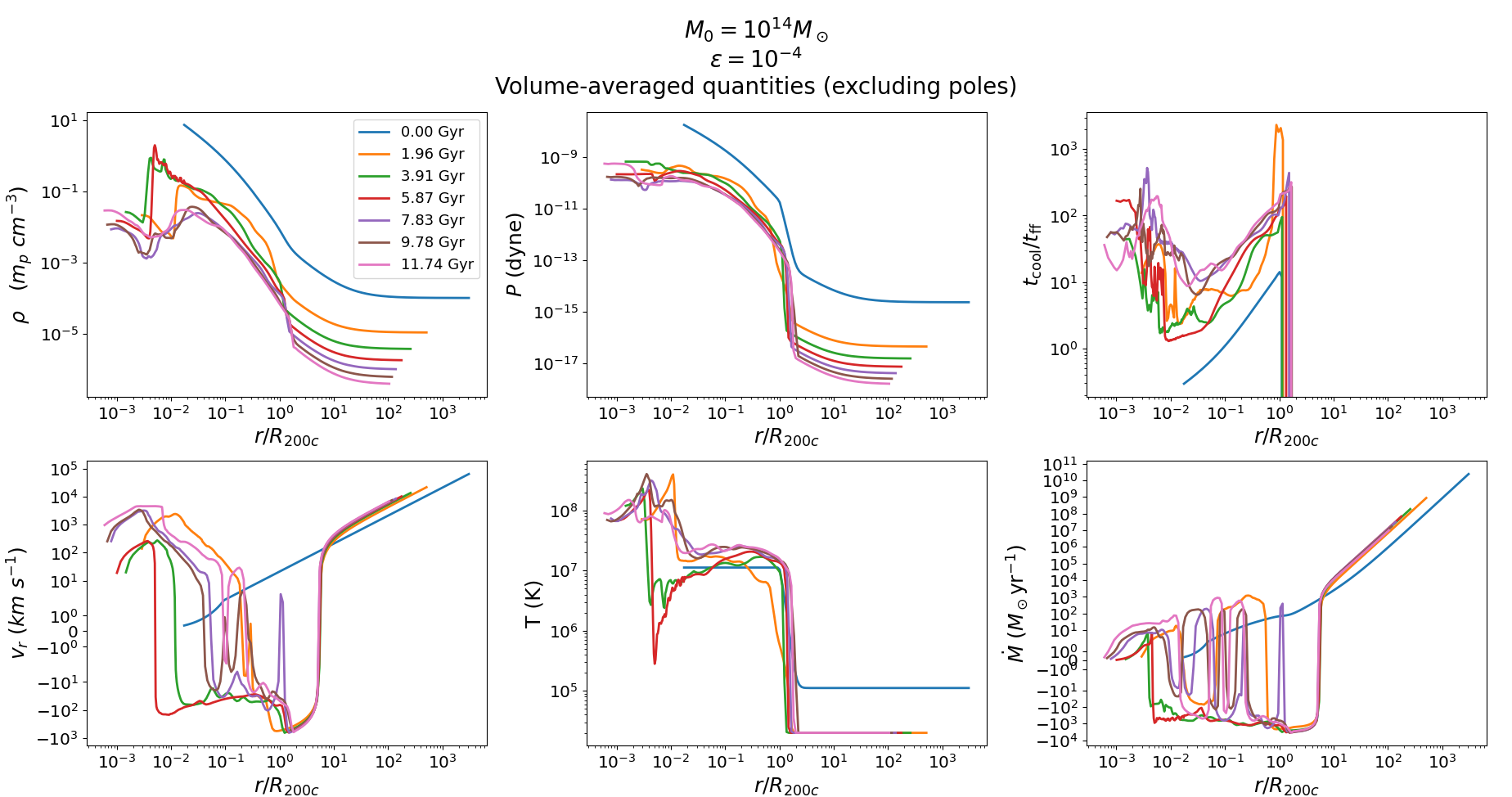}
	\caption{Various volume-averaged (excluding the poles)  profiles at different times for our fiducial AGN feedback runs; only 0.5-8  keV gas is considered for the $t_{\rm cool}/t_{\rm ff}$ profile. Feedback jets reduce the density and increases the temperature in the innermost region, while the outer regions are mostly unaffected.
	}
	\label{fig:agn_profiles}
\end{figure*}

\textit{Massive halos:} Figure \ref{fig:agn_profiles} show the various volume-averaged profiles for our fiducial feedback run ($M_0=10^{14} M_\odot$ and $\epsilon = 10^{-4}$);  $t_{\rm cool}/t_{\rm ff}$ profile only uses 0.5-8 keV gas. The jet feedback significantly increases the temperature and decreases the density in the innermost $\sim 10$ kpc of the halo compared to the corresponding cooling flow run. In the jet region, the density profile approximately follows $\rho \propto r^{-2}$ during the strong feedback phase, in agreement with the analytic solution for a wind driven by uniform mass and energy deposition \citep{cc85}. It is also clear that feedback affects only the inner regions of the halo, as the outer density, temperature, and velocity profiles are similar to the non-radiative and cooling flow runs. As we can see in the top-right panel of figure \ref{fig:agn_profiles}, the $t_{\rm cool}/t_{\rm ff}$ profile in the innermost regions is not smooth due to the presence of multiphase gas.

\begin{figure*}
	\includegraphics[scale=0.3]{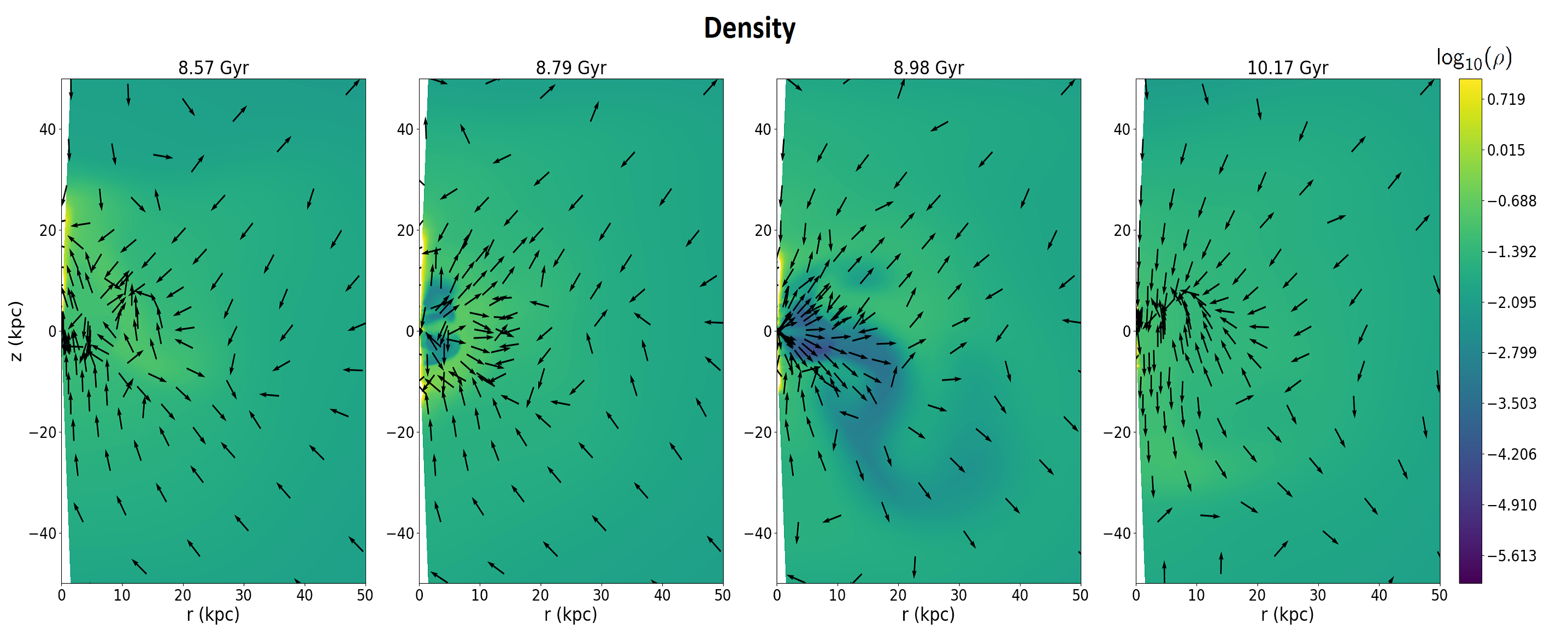}
	\includegraphics[scale=0.3]{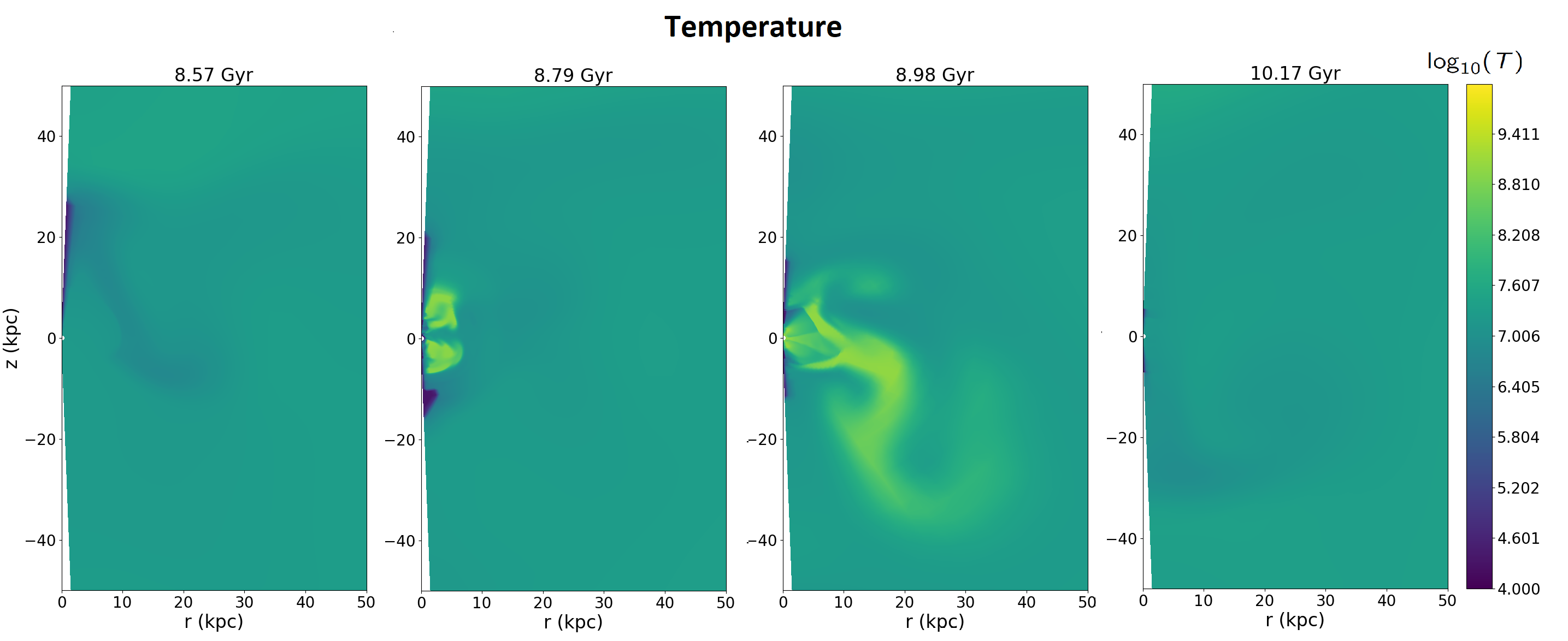}
	\caption{Snapshots of the density (top) and temperature (bottom) for our fiducial AGN feedback run ($M_0=10^{14} M_\odot$ and $\epsilon_m=10^{-4}$) over one duty cycle (see figure \ref{fig:combined_plot}). It is evident that the density and temperature are anti-correlated. The arrows in the upper plots indicate the velocity directions. Inflow of cold gas toward the center of the halo launches powerful jets which produce low density bubbles and cavities. These bubbles rise buoyantly through the ICM and mix with the surrounding medium, eventually reaching approximate pressure balance. This reduces the cold gas inflow toward the center and decreases the jet power. We note that the first and the last snapshots are similar, as they both represent a cool state of the halo core.} 
	\label{fig:snapshots}
\end{figure*}

 At very early times at the center of the halo, the gas density is high and therefore the cooling time is short. The gas rapidly cools and flows toward the center, giving rise to a high $\dot{M}_{\rm in}$. This launches the bipolar jets at the center. The jet velocity is very fast, but the jet material eventually thermalizes with the ambient CGM and reaches approximate pressure balance. The jet quickly "throws out" the central gas and drastically lowers the value of $\dot{M}_{\rm in}$, preventing further power injection into the ICM. This results in the formation of bubbles which detach from the central region and rise buoyantly. The hot, low density material of the jet mixes with the ICM, heating up the ICM core and reducing the jet power.
 
In the absence of continued energy injection from the AGN jet, the inner ICM begins to cool again. The condensation of the cold gas increases the value of $\dot{M}_{\rm acc}$ and relaunches the AGN jets. Thus, a self-regulating cycle of heating and cooling in the core is established. Figure \ref{fig:snapshots} shows the density and temperature shapshots of the fiducial AGN feedback run in the $r-z$ plane at different times over one feedback cycle. Although from our feedback prescription (equation \ref{source_dens}) the energy and momentum injection from the jets is symmetric, the evolution of the gas is not perfectly symmetric especially at later times, as it is shaped by local inhomogeneities (especially the presence of cold gas on their way).

\begin{figure}
	\includegraphics[width=1.08\columnwidth]{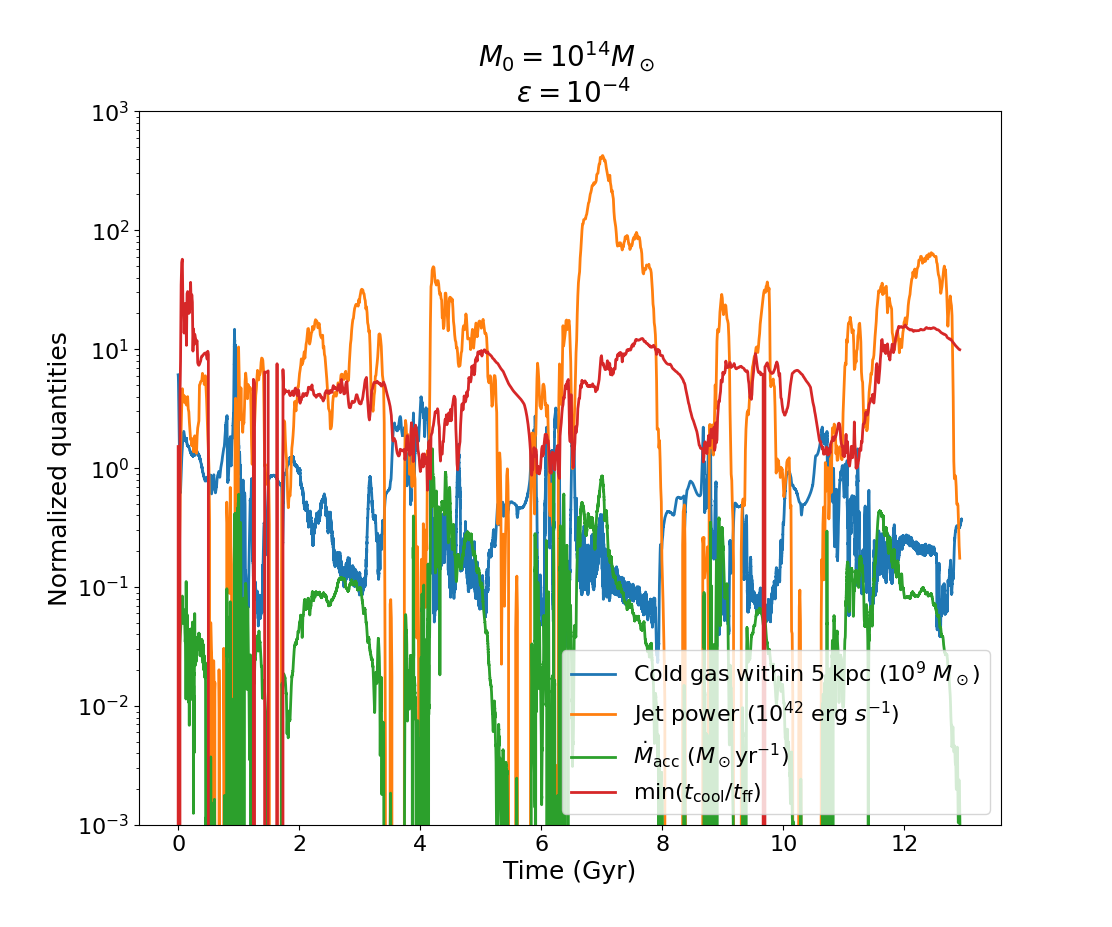}
	\caption{Normalized quantities $\dot{M}_{\rm acc}$ (accretion rate on to the SMBH), $M_{\rm cold}(<\rm 5 kpc)$ (gas mass with $T<10^5$ K within the halo), jet power (see the beginning of section \ref{sec:AGN}), and $\rm{min}(t_{\rm cool}/t_{\rm ff})$ (considering the gas between 0.5-8 keV and outside 10 kpc) for our fiducial cluster feedback run ($M_0=10^{14} M_\odot$ and $\epsilon = \epsilon_m \epsilon_{\rm BH}=10^{-4}$). The data are sampled every $\sim 10$ Myr. The spikes in jet power follow a decrease in $\rm{min}(t_{\rm cool}/t_{\rm ff})$ which triggers multiphase condensation in the core. As jet power rises (associated with the increase in BH accretion rate), the core is heated and $\rm{min}(t_{\rm cool}/t_{\rm ff})$ rises, cold gas mass within 5 kpc (a crude proxy for star formation rate) decreases, indicating heating and uplifting of gas by AGN jets. }
	\label{fig:combined_plot}
\end{figure} 
Figure \ref{fig:combined_plot} shows the cold gas mass within 5 kpc, $\dot{M}_{\rm acc}$ (the accretion rate onto the SMBH), the normalized jet power, and ${\rm min}(t_{\rm cool}/t_{\rm ff})$ as a function of time for our fiducial feedback run. We can clearly see that the values of $\dot{M}_{\rm acc}$, the normalized jet power, and ${\rm min}(t_{\rm cool}/t_{\rm ff})$ vary together. Typically, $\dot{M}_{\rm acc}$, ${\rm min}(t_{\rm cool}/t_{\rm ff})$ (because of large fluctuations within $\sim 10$ kpc, we consider the minimum value between 10 kpc and $R_{200c}$), and the jet power increase during the heating portion of a cycle, and decrease during the cooling portion of the cycle. The cold gas mass behaves in an opposite manner, decreasing in the heating portion and increasing in the cooling portion of the cycle. We expect there to be some time lag between an increase in $\dot{M}_{\rm acc}$ and an increase in the jet power and ${\rm min}(t_{\rm cool}/t_{\rm ff})$, as we first require a high accretion rate in order to launch the AGN jets that heat the gas. However, this time lag is small ($\sim 10-100$ Myr). We note that at early times in our simulations, the feedback cycles are irregular because the black hole mass (and therefore the Eddington rate) is small and the initial transients are not yet ironed out. Regular cycles are established after the initial $\sim$ 2 Gyr. 

\subsection{Varying feedback efficiency and the halo mass} \label{var_eff}
In this section, we study the evolution of the CGM with varying feedback efficiency for a range of present-day halo masses ($M_0$ varying from $10^{12}$ to $10^{15} M_\odot$). The mechanical feedback efficiency of the SMBH is fixed to be $\epsilon_{\rm BH}=0.1$ and the mass transport efficiency from $\sim 1$ kpc to the BH event horizon ($\epsilon_m$) is varied. The seed BH mass is $10^6 M_\odot$ (as we show later, the results are mostly insensitive to this choice).

The timescale of the feedback cycles depends on the cooling and heating times of the halo core, which in turn, depend on the feedback efficiency and the halo mass. AGN feedback with a high efficiency in a low-mass halo results in rapid heating of the gas and the inner material being ejected from the core, requiring a longer time for gas to cool and flow in toward the center again. Therefore, the timescale of the AGN cycles is shorter for lower efficiencies and more massive halos. 

\begin{figure}
		\includegraphics[width=\columnwidth]{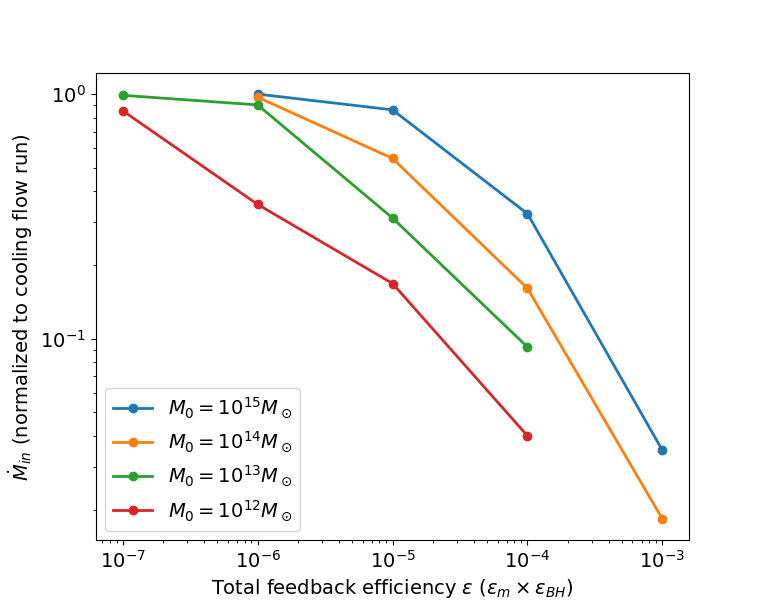}
	\hfill
		\includegraphics[width=\columnwidth]{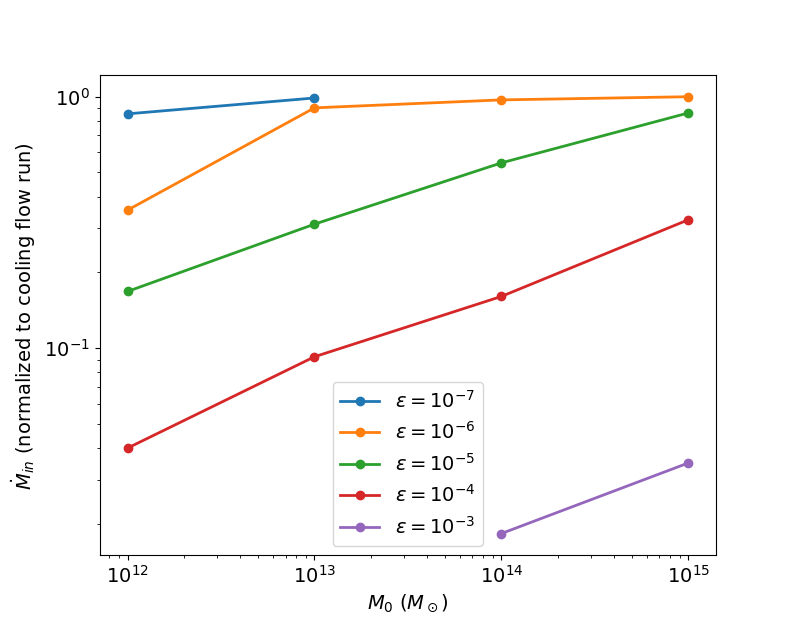}
	\caption{Variation of the time-averaged (from $z=6$ to $z=0$) mass inflow rate at $\sim$ 1 kpc with efficiency $\epsilon$ (top panel) and halo mass (bottom panel), normalized to the corresponding cooling flow run for each halo. It is clear that the inflow rate at $\sim 1$ kpc is more suppressed for greater efficiencies and lower halo masses.}
	\label{fig:mdot_in}
\end{figure}

With increase in feedback efficiency, we expect a decrease in the cold gas mass flowing in towards the core. The top panel of figure \ref{fig:mdot_in} shows the time-averaged value of the mass inflow rate at $\sim 1$ kpc ($\dot{M}_{\rm in}$) for different efficiencies, normalized to the value corresponding to the cooling flow run (no feedback). It is clear that the normalized $\dot{M}_{\rm in}$ for a particular halo decreases with $\epsilon$. Our results are consistent with \citetalias{prasad15}, who found the normalized $\dot{M}_{\rm in}$ to vary as $\sim \epsilon^{-2/3}$. 

From the bottom panel of figure \ref{fig:mdot_in}, we can see that for the same efficiency, lower mass halos show a larger suppression of the mass inflow rate at $\sim$ 1 kpc ($\dot{M}_{\rm in}$) relative to a cooling flow. This means that feedback is more effective for lower mass halos. Thus, the effect of decreasing the halo mass is similar to the effect of increasing the efficiency, consistent with \citetalias{prasad15}. Massive halos have deeper potential wells, and therefore require a greater energy in order to disrupt the core to the same degree.

\begin{figure}
	\includegraphics[width=\columnwidth]{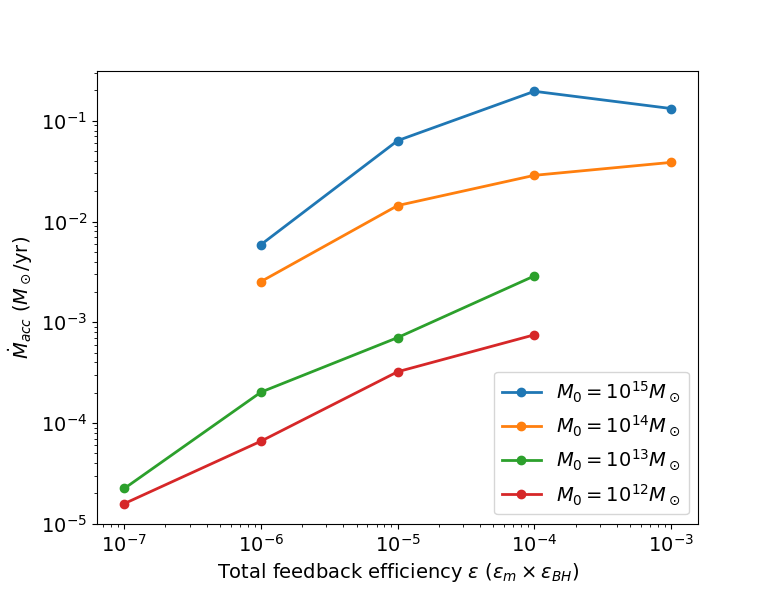}
	\caption{Variation of the SMBH accretion rate $\dot{M}_{\rm acc}$ with efficiency. The accretion rate is taken as ${\rm min}(\epsilon_m \dot{M}_{\rm in},\dot{M}_{\rm Edd})$. 	The accretion rate increases with increase in efficiency, as opposed to the inflow rate $\dot{M}_{\rm in}$ (figure \ref{fig:mdot_in}). The decrease in inflow rate at $\sim 1$ kpc due to the heating and expulsion of gas by AGN jets largely compensates the increase in the fraction of mass that is transported to the black hole event horizon.}
	\label{fig:mdot_acc}
\end{figure}

The value of the black hole mass accretion rate ($\dot{M}_{\rm acc}$) follows a different trend. Figure \ref{fig:mdot_acc} shows that the average $\dot{M}_{\rm acc}$ increases with an increase in efficiency. While this may appear counter-intuitive at first, this trend is easy to understand. Recall that the feedback efficiency $\epsilon$ is a product of two terms $\epsilon_m$ (the efficiency of mass transport from $\sim 1$ kpc to the BH event horizon) and $\epsilon_{\rm BH}$ the efficiency for the conversion of mass to mechanical energy by the central SMBH (which is held constant at 0.1). With increase in mass transport efficiency $\epsilon_m$, more gas from $\sim$ 1 kpc reaches the BH event horizon. However, there is also decrease in inflowing cold gas at the inner boundary as efficient feedback prevents gas inflow. Since $\dot{M}_{\rm in} \propto \epsilon^{-2/3}$ from figure \ref{fig:mdot_in}, we expect $\dot{M}_{\rm acc}$ (see equation \ref{eq:mdot}) to increase with efficiency as $\epsilon^{1/3}$, roughly consistent with figure \ref{fig:mdot_acc}. 

\subsection{Growth of the supermassive black hole}
\begin{figure}
	\includegraphics[width=\columnwidth]{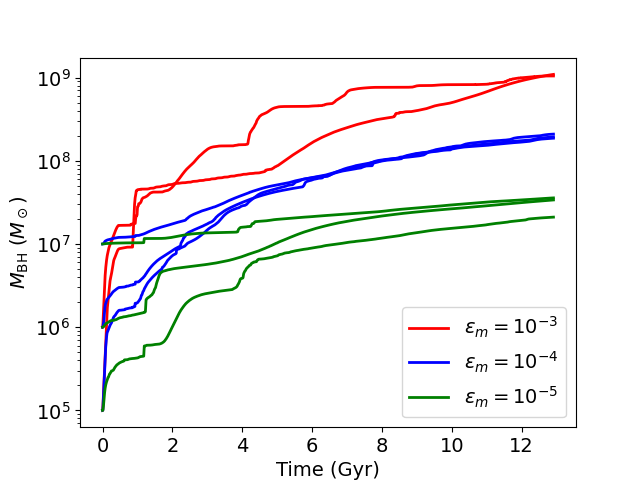}
	\caption{Growth of the central SMBH for different seed black hole masses and efficiencies for a halo mass $10^{14} M_\odot$. The black holes grow exponentially with time in the initial quasar stage of evolution. 
	It is clear that after the initial growth phase, the BH mass is largely independent of the initial seed mass, and the feedback efficiency ($\epsilon = \epsilon_{\rm BH} \epsilon_m$) plays a more important role in later evolution.}
	\label{fig:mbh}
\end{figure}

Figure \ref{fig:mbh} shows the mass of the central SMBH as a function of time for different seed black hole masses and feedback efficiencies. In the initial stages of AGN evolution, the accretion rate is limited by the Eddington rate of the low-mass SMBH. In this phase, $\dot{M}_{\rm acc}=\dot{M}_{\rm Edd} \propto M_{\rm BH}$, leading to an exponential growth of the black hole corresponding to the quasar phase of black hole evolution. As expected, the quasar phase is longer for a smaller seed black hole mass. 

After this quasar phase, we have $\dot{M}_{\rm acc} \ll \dot{M}_{\rm Edd}$ which leads to a slower growth of the black hole. The accretion rate in this phase is determined by the cooling time of the halo core and not by the Eddington limit. This leads to self-regulated growth of the central black hole independent of the seed mass, usually termed as the radio mode/phase of AGN and halo evolution. In this stage, the mass transport efficiency to the black hole event horizon ($\epsilon_m$) and the feedback efficiency of power injected into the ICM ($\epsilon_{\rm BH}$) are much more important in determining the black hole growth. 

Since the seed black hole mass only affects the initial quasar stages of evolution, we find that the seed mass is quickly forgotten by the halo, as seen in figure \ref{fig:mbh}. 
Even though the initial BH masses vary by several orders of magnitude, the mass of the black hole after $\sim 4$ Gyr mainly depends on the mass transport efficiency ($\epsilon_m$). Since the accretion rate is no longer limited by the Eddington limit in the later radio stage, we expect the evolution of the halo to be similar at later times for the same feedback efficiency, regardless of the seed black hole mass. Therefore, we conclude that the feedback efficiency ($\epsilon = \epsilon_{\rm BH}\epsilon_m$) plays a much more important role in the long-term evolution of the halo core and the SMBH than the seed black hole mass.
\subsection{Baryon fraction evolution}
\begin{figure*}
		\includegraphics[width=0.67\columnwidth]{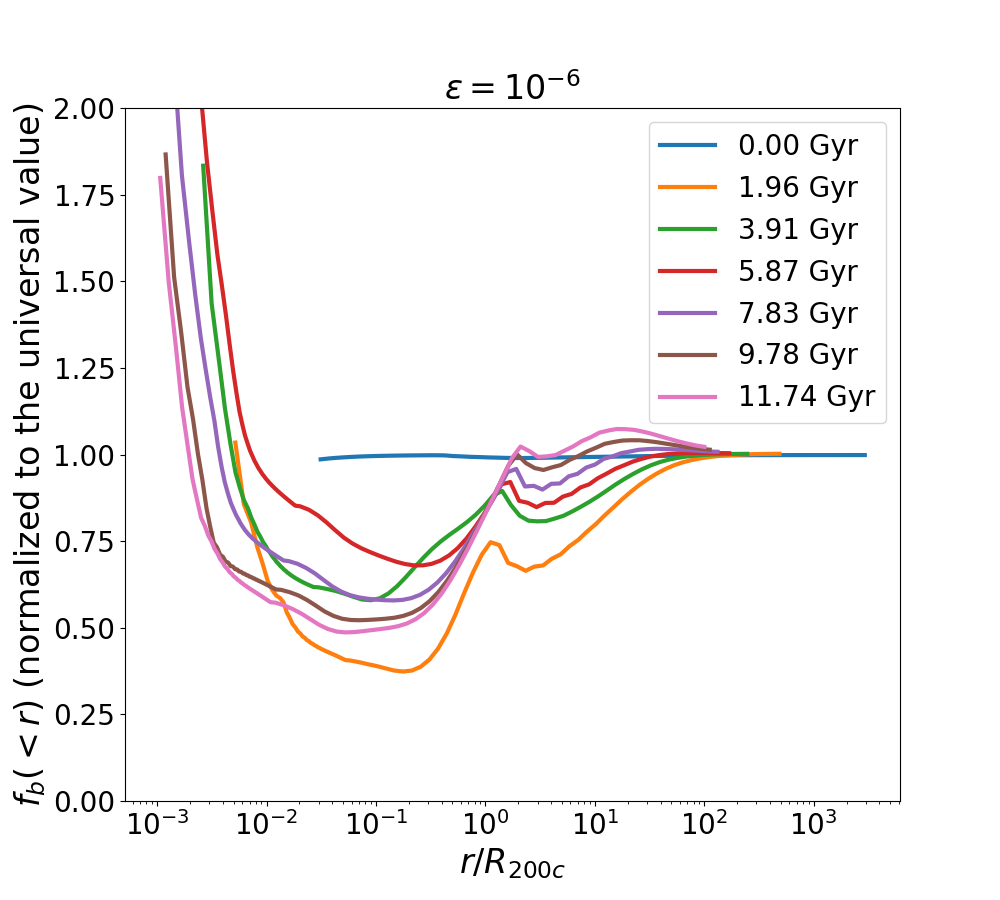}
		\includegraphics[width=0.67\columnwidth]{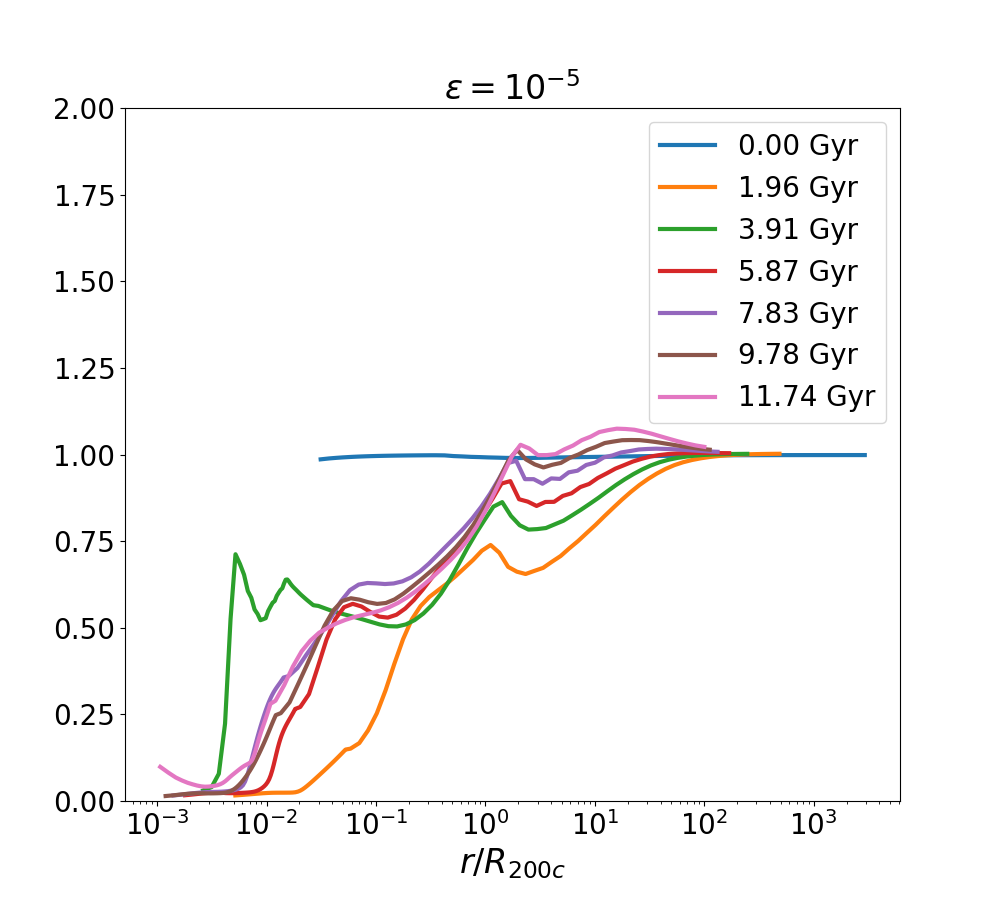}
		\includegraphics[width=0.67\columnwidth]{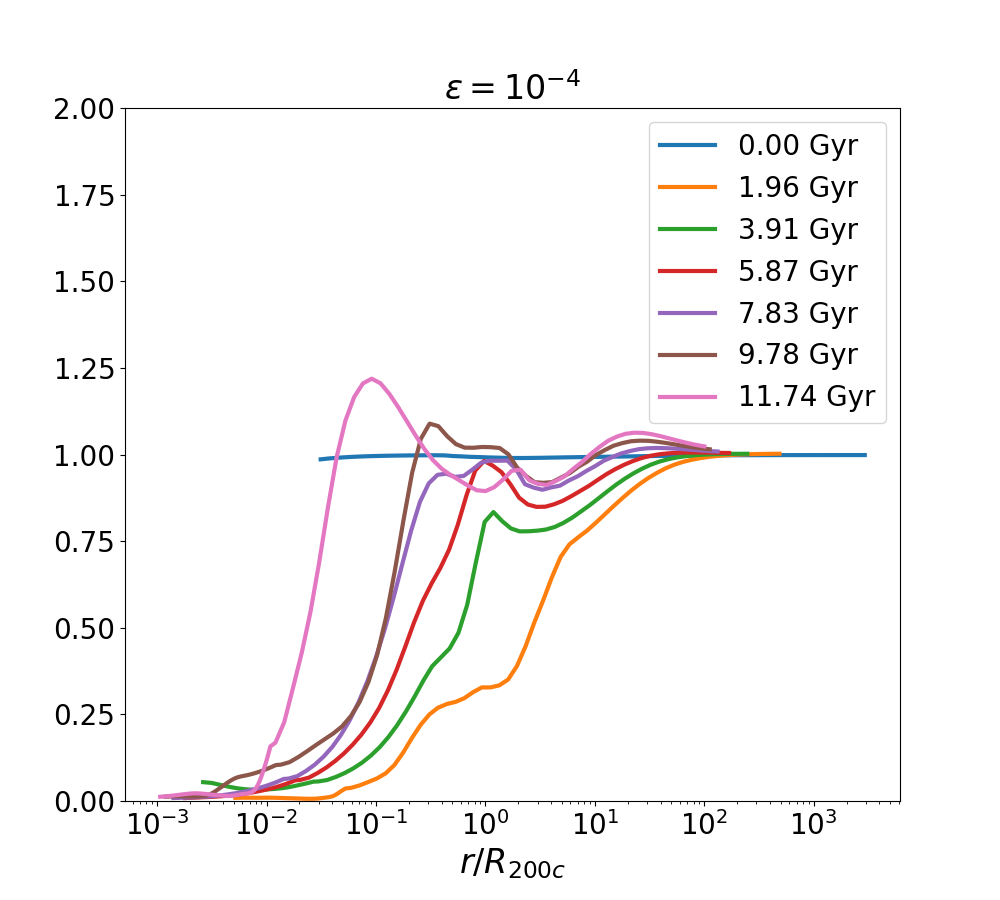}
	\caption{Normalized baryon fraction evolution for $M_0=10^{14} M_\odot$ (see table \ref{tab:dk14_parameters}) for AGN feedback efficiency $\epsilon=10^{-6}$ (left), $\epsilon=10^{-5}$ (center), and $\epsilon=10^{-4}$ (right). Similar to the effect of radiative cooling, the effects of AGN feedback are prominent only in the inner radii of the halo. The baryon fraction beyond the virial radius reaches the universal value, similar to the non-radiative and cooling flow runs. In the inner regions ($\lesssim 0.1 R_{200c}$), the baryon fraction strongly depends on the efficiency of feedback. An efficiency of $\epsilon=10^{-6}$ nearly corresponds to a cooling flow (see figure \ref{fig:mdot_in}), causing the baryon fraction within $10^{-2} R_{200c}$ to reach large values. For $\epsilon=10^{-5}$ and $10^{-4}$, the inner baryon fraction is low compared to the non-radiative runs (figure \ref{fig:new_bf}) because the AGN jet feedback leads to the formation of low-density bubbles and cavities that decrease the baryon fraction.}
	\label{fig:agn_fraction}
\end{figure*}

The jet feedback process blows AGN bubbles in the inner regions of the halo, severely decreasing the average density compared to the non-radiative and cooling flow runs (see figures \ref{fig:new_bf},\ref{fig:14cooling},\ref{fig:agn_profiles}). Therefore, the baryon fraction in the inner regions of the halo is much smaller than with non-radiative and cooling flow runs. Figure \ref{fig:agn_fraction} shows the baryon fraction as a function of radius for our $10^{14} M_\odot$ halo feedback runs for three different feedback efficiencies (parameters in table \ref{tab:dk14_parameters} are used for the outer dark matter density) 
For an easier comparison, we scale the y-axis with respect to the universal value of $1/6$.
It is clear from figure \ref{fig:agn_fraction} that feedback in cluster sized halos affects the baryon fraction in only the innermost regions of the halo, as the value of $f_b$ outside $\sim 0.1 R_{200c}$ is similar to the adiabatic and cooling flow cases (see figure \ref{fig:new_bf}), for all efficiencies.

The left panel, corresponding to $\epsilon=10^{-6}$, shows a very high baryon fraction in the core because the low feedback efficiency leads to a near cooling flow at small radii (see figure \ref{fig:mdot_in}). 
For $\epsilon=10^{-5}$ (middle panel) and $10^{-4}$ (right panel), the baryon fraction within $\sim 0.01 R_{200c}$ is at least an order of magnitude smaller than the universal value. 
We also see that the low baryon fraction region extends more outward for $\epsilon=10^{-4}$ compared to $10^{-5}$, indicating greater regulation of the core density and larger cavities in the former case.

\section{Discussion} \label{sec:discuss}
\subsection{Implementation of AGN feedback}
We use the cold mode accretion model to implement AGN jet feedback, similar to e.g. \citet{gaspari2012,li2014a}; \citetalias{prasad15}; \citet{yang2016,prasad18}. The accretion rate that \citetalias{prasad15} calculate at 1 kpc is similar to our definition of $\dot{M}_{\rm in}$. They find that a jet feedback efficiency of $6 \times 10^{-5}$ suppresses the inflow rate by a factor of 10 compared to a cooling flow in a cluster of mass $7 \times 10^{14} M_\odot$. This suppression factor is comparable to our results: we find a suppression factor $\sim 5$ for a feedback efficiency of $\epsilon=10^{-4}$ in cluster of mass $M_0=10^{15} M_\odot$. The suppression factor for a higher feedback efficiency of $10^{-3}$ is $\sim 50$, in agreement with our results (see figure \ref{fig:mdot_in}). This suppression is also in general agreement with several other previous works that do not evolve the halo cosmologically  \citep{gaspari2012,yang2016,wang2019}.

Early works of jet feedback (e.g., \citet{cattaneo2007}, \citet{dimatteo2005}, and \citet{sijacki}), particularly cosmological simulations, use the Bondi estimate to model the accretion of gas to the central black hole. Some works (e.g. \citealt{dubois2010}) include a boost factor to artificially enhance the Bondi accretion, in order to 
counter the effects of cooling (see the discussion in \citealt{valentini2021}). As noted by \citetalias{prasad15}, hot mode feedback in Bondi accretion leads to smooth increase and decrease in the accretion rate during a feedback cycle, as opposed to the rather abrupt changes in the cold mode accretion model. However, the total power injected by feedback into the ICM/CGM is more important than the specific feedback prescription \citep{li2014a}.

Although our jet implementation is similar to \citetalias{prasad15}, there are several key differences. Most importantly, we consider the cosmological evolution of the dark matter halo and metallicity evolution (albeit crude) of the IGM. The dark matter halo grows over time, leading to a deeper potential well at later times. The metallicity of the halo gas increases with time, leading to a stronger cooling at late times. These two effects result in
longer feedback cycles in our runs compared to \citetalias{prasad15}. 
We have verified that if we consider a non-evolving dark matter halo with a constant gas metallicity, our feedback cycles have a duration $\sim 100$ Myr, similar to those of \citetalias{prasad15} for a similar feedback efficiency (note that the cycle duration is longer for higher feedback efficiency). Unlike \citet{gaspari2012} and \citetalias{prasad15}, we start with small seed black holes that grow at the Eddington rate at high redshifts and in a fuel-limited radio mode at later times.  

We also make the important distinction between the mass inflow rate at $\sim 1$ kpc ($\dot{M}_{\rm in}$) and the accretion rate onto the black hole $\dot{M}_{\rm acc}$. Unlike some of the previous works on AGN feedback, we decouple the total feedback efficiency into the mass transport efficiency from 1 kpc to the black hole event horizon ($\epsilon_m$), and the efficiency of mass-to-energy conversion by the black hole ($\epsilon_{\rm BH}$). The accretion rate onto the black hole depends on both efficiencies, as the mass transport efficiency determines the amount of cold gas that can reach the black hole, and the accretion rate is limited by the Eddington rate which depends on the black hole efficiency ($\epsilon_{\rm BH}$). As examined by \citet{churazov}, the radiative efficiency of an accreting black hole depends on its accretion rate relative to the Eddington limit. We have considered a simplified model where mechanical feedback efficiency is kept a constant irrespective of the BH accretion rate (relative to the Eddington rate), but it should not affect our results as quasar phases are very short.      
\subsection{Evolution of supermassive black holes}
\label{sec:smbh_evolution}
We have implemented black hole growth in our model by using a simple prescription for the accretion of gas onto its event horizon by connecting to the cold gas accretion rate at $\sim 1$ kpc. A seed black hole is introduced at the start of the simulation ($z=6$). 
The formation of black hole seeds in the early universe can occur through several, highly uncertain, processes (a recent review is \citealt{inayoshi2020}). Using cosmological simulations, \citet{bellovary2011} find that by $z=5$, most halos with mass $M_{\rm halo}>3\times 10^9 M_\odot$ host a massive black hole. Since all halos in our simulations are more massive than this value at $z \sim 6$ (see figure \ref{fig:halo_acc}), it is therefore appropriate for us to place a massive black hole seed at the start of our simulation. Moreover, the precise value of the seed mass is not very important for long-term evolution of the halo and SMBHs (see figure \ref{fig:mbh}).

The growth of the black hole can be divided into two stages: a quasar phase with fast exponential growth, and a radio phase with slower sub-exponential growth limited by the availability of cold gas. Since the seed BHs are less massive and the CGM density is higher at higher redshift, the Eddington-limited exponential growth of the SMBH in our simulations occur at early times. This is in accordance with observational studies that most quasars occur at high redshifts \citep{vito2019}. Through cosmological simulations, \citet{hirschmann} find that the number of BHs accreting in the quasar mode is more than an order of magnitude smaller at the present redshift than at higher redshifts ($z = 1-3$). They find that quasar mode black holes are very rare at $z=0$. The black holes that exhibit quasar mode feedback at low redshifts are predominantly low mass black holes ($M_{\rm BH}<10^7 M_\odot$, see figure 14 in \citealt{hirschmann}). Observational studies \citep{vestergaard2003,kollmeier2006,kelly2010} and our SMBH evolution are consistent with this prediction. The quasar mode lasts for $\sim 100$ Myr in our simulations, although this is dependent on the seed black hole mass and the halo mass. This is in accordance with the results of \citet{sijacki2009} and \citet{volonteri2016}. 

We find that after the first $\sim 4$ Gyr, the evolution of the AGN and the gas in the halo core is largely independent of the seed black hole mass, in accordance with the results of several previous cosmological galaxy formation simulations \citep{dimatteo2009,sijacki2009, dubois2012}. The further growth is self-regulated by radiative cooling, heating, and gravitational dynamics 
in the halo core. 

AGN feedback in Milky Way-sized halos has important implications for the growth of SMBHs in these galaxies. The mass inflow rate at $\sim$ 1 kpc is lower for low mass halos than for cluster halos ($M_0 \gtrsim 10^{14} M_\odot$) because of smaller available  gas. Therefore, even for a pure cooling flow and/or high mass transport efficiency ($\epsilon_m$; see equation \ref{eq:mdot}), the mass accretion rate onto the black hole in $\sim 10^{12} M_\odot$ halo is quite low. Table \ref{tab:runs_table} shows that by $z=0$, the value of $M_{\rm BH,z=0}$ is $\sim$ few $\times 10^7 M_\odot$ for a cooling flow in a $10^{12} M_\odot$ halo, and decreases with AGN feedback. The feedback runs with $\epsilon_m \sim 10^{-4}-10^{-5}$ are able to attain a $z=0$ SMBH mass comparable to that observed in the Milky Way $\sim 4 \times 10^6 M_\odot$ \citep{gillessen2009}. We note that we use the mass transport efficiency $\epsilon_m$ as a proxy for various physical processes that occur at the center of a galaxy, including star formation, supernova feedback, and formation of cold gas clumps. In a real galaxy, all of these processes together determine the SMBH growth and star formation in the core. As an example, we note that the dark matter halo of M31 has a mass similar to the MW halo \citep{kafle}, but the central SMBH of M31 is at least an order of magnitude more massive than the central SMBH of the MW \citep{bender}. This suggests that the small-scale physics plays an important role in the SMBH evolution.

A big difference between full cosmological simulations and our work is that we consider smooth growth of halos and do not include mergers (although it should be possible to model by extending the present framework). \citet{dimatteo2009} analyze the merger tree of the earliest and most massive black holes in their simulations, and find that mergers are characterized by a short-term peak in $\dot{M}_{\rm acc}/\dot{M}_{\rm Edd}$. Thus, mergers are able to trigger short quasar events in AGN even at late times (absent in our simulations). However, \citet{lambrides2021} and \citet{sharma2021} find that AGN are not necessarily located in systems  undergoing or having undergone a major merger, suggesting that efficient cooling of the core halo gas, rather than major mergers, is more responsible for triggering quasar and AGN activity.

Since we carry out axisymmetric 2D simulations, we miss some important dynamics that occurs in 3D, such as the formation of a cold torus of $\sim$ few kpc scale (\citealt{li2014b}; \citetalias{prasad15}) (also seen in some cluster observations \citealt{david2014,mcnamara2014,russell2017}). This can affect the mass transport on to SMBHs (our variation of $\epsilon_m$ tries to understand the impact of this important missing physics).

\subsection{Evolution of halo gas}
\subsubsection{Self-similarity}
The evolution of the gas in a cluster-sized halo in the absence of cooling (figure \ref{fig:basic}) and with radiative cooling (figure \ref{fig:14cooling}) both show remarkable self-similarity outside of the innermost core. Outside the halo, the gas evolves according to the standard cosmological evolution of the mean density of the universe, i.e. $\rho \propto E(z)^2$ (see figure 3 in \citealt{mcdonald}). In addition, the density profiles in the non-radiative runs for different halo masses are reasonably self-similar with respect to the virial radius of each halo (figure \ref{fig:self_similar}). 

Self-similarity of galaxy clusters outside of the core has been confirmed by numerous observational and numerical studies. Temperature profiles of high-redshift clusters, studied using X-ray data from \textit{XMM-Newton}, have been found to follow a universal self-similar law \citep{baldi2012}. Similar results have been found from X-ray luminosity–temperature measurements from \textit{Chandra}, studying both high and low redshift clusters \citep{maughan2012, mcdonald}. Theoretical predictions of simple models applied to N-body simulations (e.g., \citealt{bohringer2012}) and SPH simulations (e.g., \citealt{battaglia2012}) are in good agreement with observed results, indicating that the physics in the outer regions of halos has been well captured. Our results on self-similarity agree with \citet{mcdonald}, who examine the self-similarity of galaxy clusters from $z \sim 0$ to $z \sim 1.9$ for $r>0.2 R_{500}$. Figure \ref{fig:self_similar} in our paper is in rough agreement with figure 8 in \citet{mcdonald}.

The virial shock of a halo in our simulations shows remarkable self-similarity at all redshifts and across a wide range of halo masses. A sudden drop in temperature and density near the virial radius has been observed in most numerical studies of baryonic gas in a dark matter halo  (e.g. \citealt{knight,dekel}). Studying the virial shock in galaxy clusters is an observational challenge because of low X-ray emissivity in cluster outskirts. However recent studies \citep{zhu2021} have found a steep decrease in the projected temperature profile near the virial radius of the Perseus cluster, which the authors have attributed to a virial shock.

\begin{figure}
	\includegraphics[width=\columnwidth]{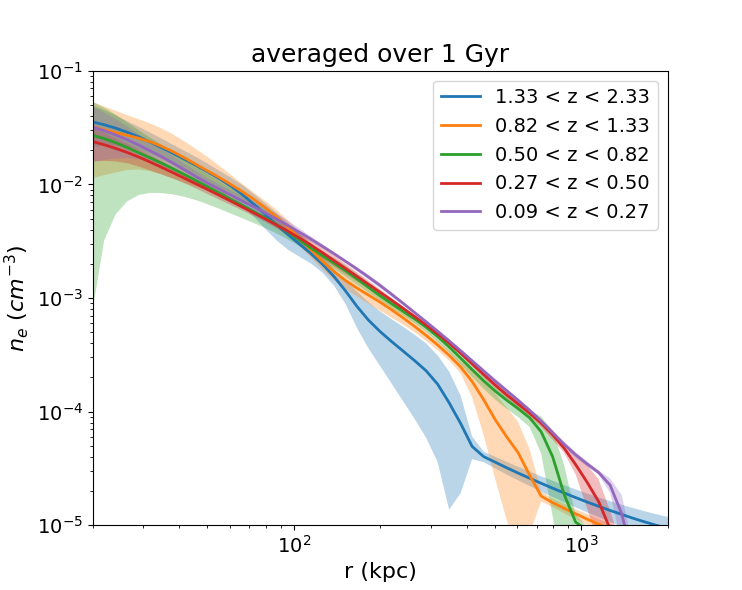}
	\caption{Time-averaged density profiles for our fiducial AGN feedback run in physical (non-scaled) units for $z \lesssim 2$. The shaded areas represent the standard deviation in each time bin. The average density in the region 20-200 kpc does not show a trend over time. On the other hand, the density in the halo outskirts is evolving self-similarly. These results are fairly consistent with figure 2 in \citet{mcdonald}. We note that our core densities are somewhat higher because we only model cool core clusters and do not include mergers.}
	\label{fig:self_similar_agn}
\end{figure}

The core of the halo, unlike the outskirts, is significantly affected by complex baryonic physics (stellar and AGN feedback, cooling, etc.), breaking the expected gravitational self-similarity. This breaking of self-similarity of cool cores is well known (see for example \citealt{ponman2003}). \citet{mcdonald} find no significant evolution of the temperature, entropy, and density in the core from $z=0$ to $z=1.9$, indicating that the properties of the core are relatively stable and non-evolving over time. They describe the cluster density as a sum of a self-similarly evolving outer profile and a non-evolving cool core. Figure \ref{fig:self_similar_agn} shows the time-averaged density profiles for our fiducial cluster feedback run ($M_0=10^{14} M_\odot$, $\epsilon=10^{-4}$). It is clear that the average density in the inner regions of the halo is non-evolving with time, whereas the outer density evolves self-similarly. This is in agreement with the results of \citet{mcdonald}.
\subsubsection{Baryon Fraction in halos}
A summary of the first generation of large-scale cosmological simulations is given in \citet{frenk1999}. They find that although the results are dependent on various factors such as cosmological parameters, resolution, and initial conditions, the average baryon fraction in most clusters reach close to the universal value near the virial radius. This was further extensively studied and verified by several authors using large-scale cosmological hydrodynamic simulations  \citep{kravtsov,crain,giguere2011,planelles,lebrun2014,diaz}. The general consensus is that the baryon fraction in the inner regions of a cluster depend on the baryonic physics (cooling increases the fraction, feedback decreases it), while the value near and outside of the virial radius is reasonably close to the universal value. This is observed in our model as well. Another important result from previous simulations is that the baryon fraction at the virial radius has a dependence on the mass of the halo. More massive halos have greater baryon fractions closer to the universal value, while lower mass halos have smaller values because a shallower potential well makes even weaker feedback more effective (see figure \ref{fig:mdot_in}). Observational studies \citep{sadat,gonzalez,chiu1,chiu2} are in agreement with this. 
\subsection{A fundamental relation between halo/SMBH properties}
\label{sec:fundamental_plane}
It has been found that the properties of a large number of elliptical galaxies lie on a relatively thin manifold, termed as the fundamental plane. \citet{faberjackson} analyzed the velocity dispersion and mass-to-light ratios of elliptical galaxies and found a correlation with the luminosity of the galaxy. Subsequent studies (e.g., \citealt{djorgovski,dressler1987,pahre1998,bernardi2003}) extended this to include various other galaxy properties like the surface brightness and effective radius. The relationship between the central black hole mass and the bulge mass of galaxies has also been studied extensively (e.g., \citealt{marconi,haring,rvdb}). A semi-analytical model for the co-evolution of black holes and their host galaxies is presented in \citet{somerville2008}, who also find that this self-regulated black hole growth naturally leads to consistent black hole-bulge mass relations. Studies (\citealt{Li2020, Habouzit2021} and references therein) have analyzed the relationships between the properties of SMBHs and the host galaxies in recent cosmological simulations. It is clear (e.g., see \citealt{Voit2020}) that black hole feedback has a direct relation with the star formation history and stellar velocity dispersion in massive galaxies. \\
\begin{figure}
	\includegraphics[width=\columnwidth]{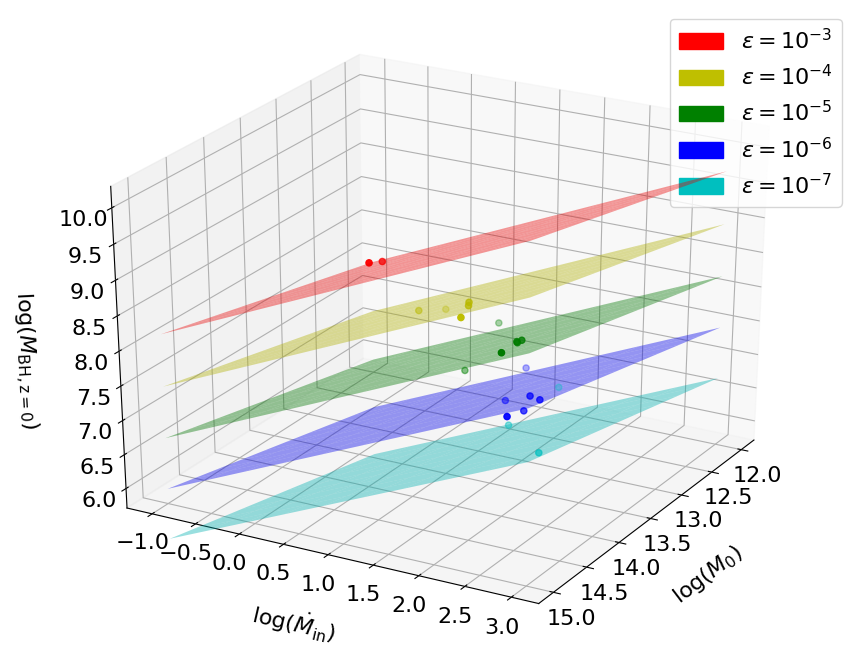}
	\caption{Properties of the halo gas and central SMBH in the $\log(M_0)$ - $\log(\dot{M}_{\rm in})$ - $\log(M_{\rm BH,z=0})$ space for different feedback efficiencies. These quantities approximately lie on a manifold defined by equation \ref{eqt:funda_plane}. It is clear that more massive black holes are found in massive halos, halos with a high $\dot{M}_{\rm in}$ (corresponding to a larger star formation rate) and halos with a greater feedback efficiency (specifically, more efficient transport of gas from $\sim 1$ kpc to the event horizon). The reader can download the Python code from the supplementary material and visualize this figure from different angles.}
	\label{fig:fundamental_plane}
\end{figure}
We select the present-day black hole mass $M_{\rm BH,z=0}$, the present-day halo mass $M_0$, time-averaged cold gas inflow rate $\dot{M}_{\rm in}$, and the feedback efficiency $\epsilon$ as fundamental quantities and check for correlations among these. Note that we do not include the seed black hole mass as a fundamental quantity because, as described in section \ref{sec:smbh_evolution}, the present-day BH mass and the halo gas properties are mostly independent of the seed BH mass. We fit these four quantities to the equation
\begin{equation}
\label{eqt:funda_plane}
    M_{\rm BH,z=0} = A \ M_0^B \ \dot{M}_{\rm in}^C \ \epsilon^D.
\end{equation}
The best-fit coefficients for the above equation (obtained from fitting a linear relation to the logarithm of equation \ref{eqt:funda_plane}) are:
\begin{equation}
\label{eq:best_fit_plane_params}
\begin{split}
    &A=7.15 \pm 1.77,\\
&B=0.26 \pm 0.12, \\ 
&C=0.58 \pm 0.15, \\
&D=0.75 \pm 0.07.\\
\end{split}
\end{equation}
These quantities have strong correlations with each other. 

Figure \ref{fig:fundamental_plane} shows $\log(M_{\rm BH,z=0})$ for different runs and the best-fit planes in the $\log(M_0)$ - $\log(\dot{M}_{\rm in})$ - $\log(M_{\rm BH,z=0})$ space for different feedback efficiencies. It is clear that the points lie in a relatively small region in the total space defined by these four variables. Thus, the black hole mass is well correlated with all three of the other quantities. The positive correlation between present-day BH mass ($M_{\rm BH,z=0}$) and halo mass ($M_0$) quantitatively shows that more massive black holes are found in more massive halos, as discussed in section \ref{sec:smbh_evolution}. We also note that $M_{\rm BH,z=0}$ depends the most strongly on the feedback efficiency $\epsilon$, which highlights the importance of AGN feedback in the cosmological evolution of SMBHs. 

If the three quantities $M_0$, $\dot{M}_{\rm in}$, and $\epsilon$ are all taken as independent variables, equation \ref{eq:best_fit_plane_params} predicts a positive correlation between the black hole mass and the mass inflow rate when $M_0$ and $\epsilon$ are fixed. However, figure \ref{fig:mdot_in} shows that for a fixed halo mass, $\dot{M}_{\rm in}$ decreases with increase in $\epsilon$. If we account for this dependency and hold $M_0$ fixed, the black hole mass is then negatively correlated with $\dot{M}_{\rm in}$, more specifically using equation \ref{eq:best_fit_plane_params} and $\dot{M}_{\rm in} \propto \epsilon^{-2/3}$,
\begin{equation}
    M_{\rm BH,z=0} \propto \dot{M}_{\rm in}^{-0.55}.
\end{equation}
The cold gas mass inflow rate $\dot{M}_{\rm in}$ can be taken as an indirect measure of the star formation rate (SFR) in the central galaxy. Our analysis therefore shows that a lower SFR is expected in halos with more massive SMBHs, which in turn are present in more massive halos. This is in agreement with observations (\citealt{terrazas2017}) and simulations (\citealt{terrazas2020}), which find a decrease in the specific SFR with $M_{\rm BH}$. 

\section{Conclusions} \label{sec:conc}
We have studied the evolution of baryonic gas in  isolated dark matter halos. As opposed to other idealized studies that use an NFW profile to characterize the halo, we use a profile that transitions from the NFW profile within halos to a more realistic flatter profile outside \citepalias{dk14}. This allows us to study the evolution of the gas beyond the virial radius of the halo, while simultaneously analyzing the effects of cooling and feedback in the halo core. We have modeled important physical mechanisms that occur inside dark matter halos in an idealized setup, including cosmological growth of the halo, Hubble expansion at large radii, radiative cooling, AGN jet feedback, evolution of the IGM metallicity, and growth of the central supermassive black hole (including the initial quasar mode). 

The important conclusions of our study are as follows:
\begin{enumerate}
    \item In the absence of radiative cooling and feedback, the density profiles of the gas behave self-similarly after  initial transients. The density profiles have a core with a nearly constant temperature that depends on the halo mass. It is important to model the dark matter density beyond the virial radius because the accretion rate onto the halo depends on the density/gravity outside of the halo. Using redshift-dependent parameters to characterize the DM density far from the halo, we find that the baryon fraction at the virial radius is within $\sim 20 \%$ of the universal baryon fraction at all times.
    \item  Radiative cooling in low-mass halos leads to catastrophic cooling within the halo, and no virial shock is formed. On the other hand, for high-mass halos, radiative cooling significantly affects only the inner regions of the halo, where the baryon fraction is much higher than the universal value.
    \item The presence of AGN feedback leads to cooling and heating cycles in the halo core, similar to the results of \citetalias{prasad15}. Typically, the heating portion of the feedback cycle is characterized by greater jet power, enhanced accretion onto the central SMBH, and decrease in the cold gas mass in the inner regions. This is also reflected by an increase in ${\rm min}(t_{\rm cool}/t_{\rm ff})$. 
    \item A greater feedback efficiency and a smaller halo mass both lead to stronger feedback events and a decrease in mass inflow rate at $\sim$ 1 kpc. On the other hand, the gas accretion onto the SMBH increases with efficiency of mass transport from $\sim$ 1 kpc to the black hole event horizon. For $M_0=10^{14} M_\odot$, we find that an efficiency of $\epsilon \sim 10^{-4}$ (relative to the mass inflow rate measured at $\sim$ 1 kpc) suppresses the mass inflow rate by a factor of $\approx 6$ compared to the cooling flow run, which is roughly consistent with observations and previous results.
    \item The growth of the central SMBH can be divided into two regimes: a short quasar period where the accretion rate is high ($\dot{M}_{\rm acc} \sim \dot{M}_{\rm Edd}$) and a longer radio period where the accretion rate is low ($\dot{M}_{\rm acc} \ll \dot{M}_{\rm Edd}$). The duration of the quasar phase, characterized by an exponential increase in SMBH mass, primarily depends on the seed black hole mass and the efficiency of mass transport to the event horizon, and typically lasts for $10-100$ Myr. At later stages, the black hole growth slows down and is mostly independent of the seed black hole mass. The halo mass and feedback efficiency play a more important role in the further evolution of the AGN.
    \item The present-day black hole mass, present-day halo mass, mass inflow rate at 1 kpc, and the feedback efficiency of a halo are tightly correlated. These quantities lie on a plane in the $\log(M_{\rm BH,z=0})$ - $\log(M_0)$ - $\log(\dot{M}_{\rm in})$ - $\epsilon$ space, as discussed in section \ref{sec:fundamental_plane}. 
\end{enumerate}
Our framework for studying gas in isolated dark matter halos  leads to results that are consistent with previous studies, and extends them to higher redshifts. A middle ground between full cosmological simulations and idealized boxes, we hope that our model will provide a useful framework to understand the evolution of the CGM across redshifts and halo masses. 

\section{Acknowledgements}
We thank the anonymous referee for the useful suggestions that greatly improved the quality of the paper. We thank Prakriti Pal Choudhury and Benedikt Diemer for helpful discussions. PS acknowledges a Swarnajayanti Fellowship (DST/SJF/PSA-03/2016-17) and a National Supercomputing Mission (NSM) grant from the Department of Science and Technology, India. 

\section{Data availability}
We have hosted some of the codes and data files used in our work in this \href{https://github.com/shashankd98/Semi_Cosmo}{Github repository} for public access. The full data associated with this work will be shared on a reasonable request to the authors.  

\bibliographystyle{mnras}
\bibliography{references}

\bsp	
\label{lastpage}
\end{document}